\documentclass[12pt]{article} 

\usepackage{newtxtext,newtxmath}

\usepackage{graphicx}
\usepackage[normalem]{ulem}
\usepackage{changes}

\usepackage{comment}
\usepackage[letterpaper,margin=1in]{geometry}

\renewenvironment{abstract}
	{\quotation}
	{\endquotation}

\date{}

\makeatletter
\renewcommand{\fnum@figure}{\textbf{Figure \thefigure}}
\renewcommand{\fnum@table}{\textbf{Table \thetable}}
\makeatother

\usepackage{scicite}          

\def\scititle{Photonic time crystals assisted by quasi-bound states in the continuum}
\title{\bfseries \boldmath \scititle}

\author{
	Puneet Garg$^{1\ast}$,
	Evangelos Almpanis$^{2,3}$,
	Leander Zimmer$^{1}$,
    Jan David Fischbach$^{4}$,\and
    Xuchen Wang$^{5}$,
    Mohammad S. Mirmoosa$^{6}$,
    Markus Nyman$^{1}$,
    Nikolaos Stefanou$^{2}$,\and
    Nikolaos Papanikolaou$^{3}$,
    Viktar Asadchy$^{7\ast}$,
    Carsten Rockstuhl$^{1,4\ast}$\and
	\small$^{1}$Institute of Theoretical Solid State Physics, Karlsruhe Institute of Technology, Kaiserstrasse 12,\and \small Karlsruhe, 76131, Germany\and
	\small$^{2}$Section of Condensed Matter Physics, National and Kapodistrian University of Athens, Panepistimioupolis,\and \small Athens, 15784, Greece\and
    \small$^{3}$Institute of Nanoscience and Nanotechnology, NCSR “Demokritos", Patriarchou Gregoriou and Neapoleos Str.,\and \small Ag. Paraskevi, Athens, 15310, Greece\and
    \small$^{4}$Institute of Nanotechnology, Karlsruhe Institute of Technology, Kaiserstrasse 12, Karlsruhe, 76131,\and \small Germany\and
    \small$^{5}$College of Physics and Optoelectronic Engineering, Harbin Engineering University, Nantong Street 145,\and \small Qingdao, 150001, China\and
    \small$^{6}$Department of Physics and Mathematics, University of Eastern Finland, Yliopistokatu 7, Joensuu, 80101,\and \small Finland\and
    \small$^{7}$Department of Electronics and Nanoengineering, Aalto University, Maarintie 8, Espoo, 02150,\and \small Finland\and
	\small$^\ast$Corresponding authors. Email: puneet.garg@kit.edu, viktar.asadchy@aalto.fi, carsten.rockstuhl@kit.edu 
}

\begin{document} 

\maketitle
\paragraph*{Teaser:} High-Q quasi-bound states in the continuum reduce the modulation amplitude required for photonic time crystals.

\paragraph*{Short Title:} Photonic time crystals assisted by qBICs.

\newpage
\begin{abstract} \bfseries \boldmath
Photonic time crystals (PTCs) are characterized by the rapid modulation of the material properties in time, causing a momentum bandgap for light. However, the observation of these bandgaps at optical frequencies remains elusive, as the necessary temporal modulation amplitudes to show notable momentum bandgaps are relatively high, inaccessible with available materials. While it has been known that structuring PTCs at the sub-wavelength scale can improve the bandgap size, we push this concept to the extreme by leveraging the nanophotonic toolbox. Specifically, we demonstrate that structures composed of scatterers supporting quasi-bound states in the continuum can substantially reduce the required modulation amplitudes by enhancing the interaction time between light and time-varying matter. This allows us to observe noticeable momentum bandgaps despite the weak temporal modulation. Our approach bridges the concepts of bound states in the continuum and time-varying metamaterials, paving the way toward realizable PTCs at optical frequencies.
\end{abstract}

\section*{Introduction}
\noindent
Recently, there has been a surge of interest in the physics of time-varying photonic media \cite{galiffi2022photonics,hayran2021capturing,Yang2024ultrafast,schiller2024time,engheta2023four,Engheta2021metamaterials,liberal2025synthetic,prudencio2023synthetic,huidobro2021homogenization, galiffi2026optical,koufidis2024electromagnetic,pacheco-peña2020antireflection,mirmoosa2025quantum}. In particular, photonic time crystals (PTCs), \textit{i.e.}, materials characterized by periodic modulation of the permittivity in time, have been at the forefront of this research domain \cite{Cassedy1962temporal,zurita-sanchez2009reflection,asgari2024theory, Koutserimpas2022parametric,Rizza2022nonlocal,ramaccia2020phase}.  
The salient property of PTCs is that they exhibit momentum bandgaps. Inside these bandgaps, eigenmodes exist with complex frequencies for a certain range of wavenumbers. Depending on the sign of the imaginary part, the complex frequency modes amplify the incident radiation that couples to them \cite{zurita-sanchez2009reflection,zhang2024longitudinal}. 
Photonic time crystals show promise for intensifying light-matter interactions, much like spatial photonic crystals enhance light confinement~\cite{john1987strong,yablonovitch1987inhibited}. Applications of momentum bandgaps in PTCs include amplified spontaneous emission~\cite{lyubarov2022amplified,park2025spontaneous}, subluminal Cherenkov radiation~\cite{dikopoltsev2022light}, and temporal localization effects~\cite{carminati2021universal,sharabi2021disordered,apffel2022experimental, coppolaro2025exploring}, among many others.

Although PTCs have been recently experimentally demonstrated at microwave frequencies thanks to the availability of fast variable electronic and radio frequency (RF) components~\cite{reyes-ayona2015observation,park2022revealing,wang2023metasurface}, their realization at even higher frequencies remains a major challenge. Achieving the required temporal modulation demands extremely fast material changes -- typically at twice the oscillation frequency of the probing light -- and a relative permittivity change close to unity when considering a bulk (\textit{i.e.}, a spatially homogeneous) PTC~\cite{zurita-sanchez2009reflection,hayran2022dispersion}. 
The optical modulation of transparent conducting oxides (TCOs) by an external pump has emerged as a viable path to study effects related to time-varying media. In particular, indium tin oxide (ITO) and aluminum-doped zinc oxide (AZO) offer relative permittivity changes of up to 100\% in their epsilon-near-zero (ENZ) frequency regime~\cite{alam2016large,bohn2021spatiotemporal,tirole2023double,zhou2020broadband,Jaffray2025spatio}. The permittivity modulation in TCOs can be achieved using hot electron nonlinearities \cite{Jaffray2025spatio} as well as third-order nonlinearities \cite{galiffi2026optical}. However, achieving such modulation requires extremely high pumping power densities~\cite{hayran2022dispersion}. Combined with substantial material dissipation, these requirements pose a risk of thermal damage, making practical implementation of PTCs with TCOs highly challenging~\cite{lustig2023time}. Moreover, achieving a ``pure PTC" regime using time-varying bulk TCO platforms may prove difficult ~\cite{khurgin2024photonic,hayran2022dispersion}.

A loophole to achieve a PTC not obscured by other effects is to use third-order nonlinearity in transparent materials, such as Si, Ge, or GaAs~\cite{hayran2022dispersion, Hon2011the}. For example, the process of third-order nonlinearity (under the assumption of an undepleted pump) can be modeled using linear coupled-mode equations of a spatially uniform material with time-varying permittivity. The modulation frequency in this case can be extremely high, virtually unlimited~\cite{hayran2022dispersion}. However, in low-loss materials, nonlinear effects are inherently weak~\cite{borchers2012saturation}, restricting the relative permittivity change to a maximum of not more than 1\%. This tiny modulation amplitude is insufficient to observe the main features of PTCs. Indeed, inevitable scattering losses in realistic material systems and other imperfections may easily obstruct the weak signal amplification.
Nevertheless, a recent theoretical study demonstrated that momentum bandgaps in low-loss PTCs can be expanded even with weak permittivity modulation amplitudes by leveraging structural resonances in a metasurface~\cite{wang2025expanding}. The proposed metasurface comprised an array of subwavelength resonant silicon spheres that support dipolar Mie resonances. However, this design suffers from limitations in that the resonances had low quality factors (Q-factors), which necessitate an operation close to the laser-induced damage threshold of realistic materials~\cite{malik2013determination,Austin2018femtosecond,kern2011comparison, grinblat2017efficient}.

To overcome this limitation, we leverage here the concept of photonic quasi-bound states in the continuum (qBICs) and combine it with PTCs. Photonic qBICs are known to exhibit arbitrarily high Q-factors, limited only by material losses, finite-size effects, and fabrication tolerances~\cite{hsu2016bound,rybin2017high,koshelev2020subwavelength,bognadov2019bound}. We demonstrate that metasurfaces made of nanoresonators supporting qBICs pave the way for the realization of practical PTCs using conventional materials and accounting for realistic losses. Specifically, we propose two metasurface geometries based on multilayered spherical and homogeneous cylindrical nanoresonators. The study of multilayered sphere-based metasurfaces sheds light on the theoretical aspects of expanding momentum bandgaps with qBICs. In particular, with this geometry, we study the effects of \textit{material-induced} qBICs. Meanwhile, the cylinder-based metasurface demonstrates that noticeable momentum bandgaps are achievable despite realistic losses with reasonable designs. We show that these bandgaps are achieved while operating at pump powers an order of magnitude below the laser-induced damage threshold. The cylinder-based metasurface utilizes \textit{geometry-induced} qBICs. Therefore, for completeness, both structures are discussed. We also highlight that our work extends the Mie theory framework to time-varying metasurfaces made from multilayered spheres and finite-sized cylinders \cite{waterman1965proceedings}.

\begin{figure*}
\centerline{\includegraphics[width= 1\columnwidth,trim=0.1 0.1 0.1 1,clip]{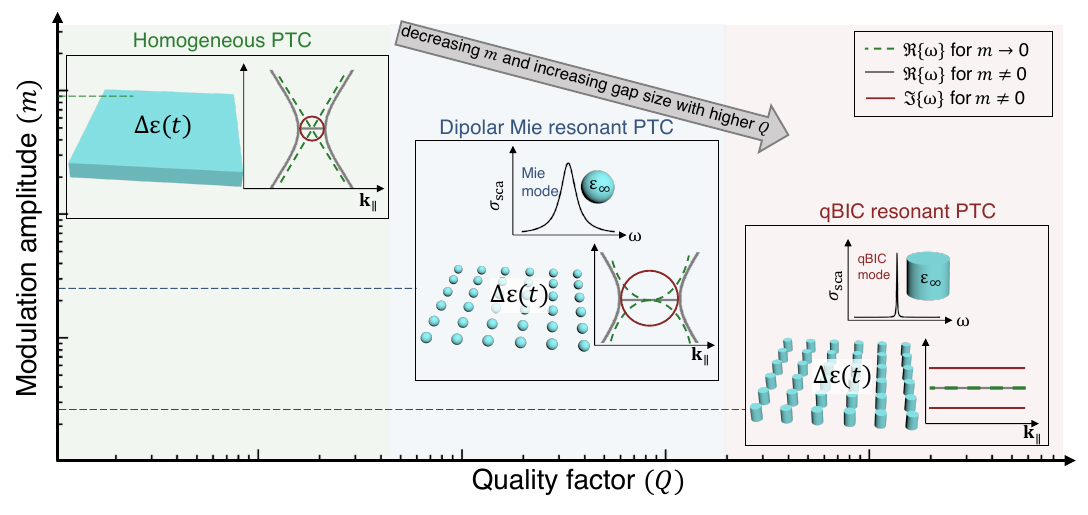}}
\caption{\textbf{A conceptual illustration}. This schematic discusses the modulation amplitude $m$ of the material properties necessary to open momentum bandgaps as a function of the quality factor $Q$ of the resonances supported by meta-atoms in various PTC geometries. We observe that, as $Q$ increases, $m$ decreases for increasing gap size. The illustration depicts a spatially homogeneous PTC (left column), a PTC constructed from meta-atoms that sustain the lowest-order Mie resonances (middle column), and the proposed PTC whose meta-atoms host high-Q qBICs (right column). The Mie and qBIC resonances of the static meta-atoms having background permittivity $\varepsilon_\infty$ are depicted in the plots of their respective scattering cross sections $\sigma_\mathrm{sca}$. Here, $\Delta\varepsilon(t)$ represents the time-periodic background permittivity of the underlying media of the meta-atoms, $\omega$ is the frequency, and $\mathbf{k}_\parallel$ is the in-plane wave vector of the eigenmodes of the PTCs. Besides the basic illustration of the geometry and resonances, the band structure is shown for zero (dotted green) and non-zero (solid gray and solid red) modulation amplitude $m$. Note that the high-Q resonance of an isolated static meta-atom leads to a strong localization of the field inside the meta-atom. This localization reduces the interaction between the neighboring meta-atoms when they are placed in a lattice, leading to reduced spatial dispersion in the lattice. Therefore, a flat band emerges in the band structure of the corresponding static metasurface. 
}
\label{fig:concept}
\end{figure*}

\section*{Results}
\subsection*{Concept}
We illustrate the proposed physical idea in Fig.~\ref{fig:concept}. A conventional PTC is spatially homogeneous and consists of a medium whose material properties, such as the background permittivity $\Delta\varepsilon$, change periodically in time, \textit{i.e.}, $\Delta \varepsilon(t)=\varepsilon_\infty\left\{1+m\right[1+\mathrm{cos}(\omega_\mathrm{m}t)\left]\right\}$. Here, $\varepsilon_\infty$ is the background permittivity of the corresponding static medium, $m$ is the modulation amplitude, and $\omega_\mathrm{m}$ is the modulation frequency. Note that the background permittivity refers to the effective non-resonant contribution to total permittivity (see Supplementary Section~\ref*{supp-sec:kerr}) \cite{galiffi2026optical}. We use the optical Kerr effect to account for the time-varying permittivity~\cite{khurgin2024photonic,xiao2014propagation,galiffi2026optical}. In this scenario, the parameters of the permittivity modulation can be expressed as $m=\frac{\chi^{(3)}E_\mathrm{p}^2}{2\varepsilon_\infty}$ and $\omega_\mathrm{m}=2\omega_\mathrm{p}$ (see Supplementary Sections~\ref*{supp-sec:kerr} and \ref*{supp-sec:homogeneous-pump}). Here, $\chi^{(3)}$ is the third-order nonlinear susceptibility of the medium, and $E_\mathrm{p}$ and $\omega_\mathrm{p}$ are the real-valued field amplitude and the frequency of the pump wave, respectively. The dispersion relation of light in the bulk static medium in its low material loss regime is linear. The periodic time modulation, in a temporal empty lattice approximation (\textit{i.e.}, $m\rightarrow0$ and $\omega_\mathrm{m}\neq0$), causes the periodic repetition of the dispersion relation in the frequency domain. Each branch is displaced by a multiple of $\omega_\mathrm{m}$. These are the Floquet bands, and they cross at the edges of the frequency Brillouin zone. When switching on a finite amplitude for the time modulation (\textit{i.e.}, $m\neq0$), a momentum bandgap opens around the crossing point. Within the momentum bandgap, the eigenfrequencies can have positive or negative imaginary parts. Assuming a time convention of $e^{-i\omega t}$, these two solutions exponentially grow or decay in time, respectively, upon excitation. However, the achievable momentum bandgap remains rather tiny, even for substantial modulation amplitudes \cite{hayran2022dispersion}.

To overcome this limitation, we can spatially structure the PTC and exploit basic photonic resonances supported by the structured material. A prototypical example of a photonic resonance is that of a Mie resonance. It occurs whenever a multipole moment sustained in the spatially localized scatterer is resonantly excited. Due to such a resonance, light remains trapped inside the scatterer for an extended duration, where it experiences the time-varying material properties. 

When periodically arranging such Mie scatterers to form a metasurface, the bands in the static material experience a rather strong curvature as a consequence of the resonant modulation of the otherwise linear dispersion relation (see Fig.~\ref{fig:concept}). Switching on the time modulation with a modulation frequency approximately twice the resonance frequency causes a repetition of the dispersion relation in the frequency domain, a crossing at the spectral location of the resonance, and a larger momentum bandgap. However, the approach has its limitations. The achievable Q-factors for ordinary Mie resonances remain rather low for the lowest-order multipole moments. These resonances couple efficiently to free space and therefore suffer from high radiation losses.

However, nanophotonics provides us with the tools to perceive scatterers with little to no radiation losses. These structures host bound states in the continuum (BICs) and are designed to have entirely suppressed radiative losses \cite{monticone2014embedded,koshelev2020subwavelength,rybin2017high}. Admittedly, the complete suppression of radiation losses is not desirable, as we are still required to excite these resonances. Nevertheless, by deliberately breaking the symmetry or detuning the system incrementally from the BIC condition, we can deterministically adjust the Q-factor that the scatterer sustains. The resulting high-Q resonances are qBICs.

When placing these scatterers with qBICs in a metasurface lattice, a photonic flat band emerges for the stationary medium (see Fig.~\ref{fig:concept}). Now, by choosing $\omega_\mathrm{m}$ to be approximately twice the qBIC resonance frequency, in theory, an extremely wide momentum bandgap emerges for a tiny modulation amplitude. The crossing of the two flat bands, split by time modulation, eventually covers the entire Brillouin zone. By using our understanding of how to structure materials to induce resonances with ultrahigh Q-factors, we can pave the way for the realistic implementation of PTCs and their associated momentum bandgaps.

It is important to note that, in this work, we exploit the flat bands of the metasurfaces to expand momentum bandgaps. The flatness of the bands is predominantly controlled by the Q-factors of the resonances of the isolated meta-atoms. Therefore, we optimize the individual meta-atoms (and not the metasurface) to host high-Q qBICs (see Supplementary Section~\ref*{supp-sec:lattice-bic}).

In the following, we illustrate this concept with two different examples that achieve such scatterers with high Q-factors by leveraging different nanophotonic approaches. The first example concerns a metasurface made from ENZ-material-based multilayered spheres. The second example concerns a metasurface composed of germanium cylinders. In both examples, we capitalize on high-Q qBICs to show that the modulation amplitude required to obtain noticeable momentum bandgaps can be considerably reduced as compared to homogeneous PTCs and metasurfaces supporting ordinary Mie resonances. 

\subsection*{PTCs based on metasurfaces made from multilayered spheres}
\begin{figure*}
\centerline{\includegraphics[width= 1\columnwidth,trim=0.1 0.1 0.1 0.1,clip]{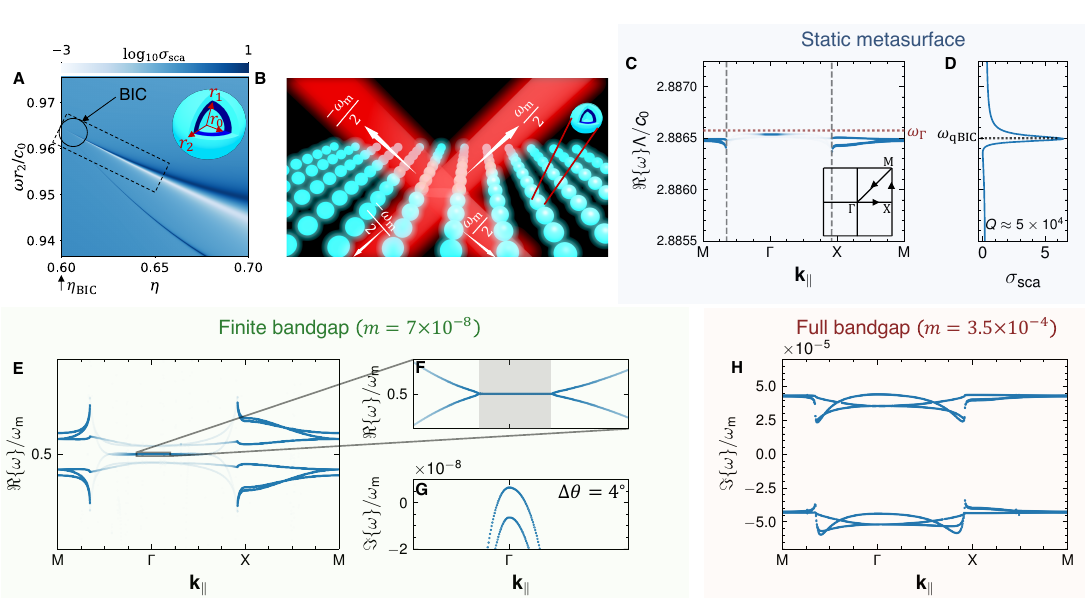}}
\caption{\textbf{PTC based on multilayered sphere metasurface.} (\textbf{A}) The normalized scattering cross section $\sigma_\mathrm{sca}$ of the static multilayered sphere (see inset) as a function of the frequency $\omega$ and the geometry parameter $\eta=r_1/r_2$. The black circle marks the spectral location of the true BIC. Note that material losses are neglected here to clearly illustrate the effects. (\textbf{B}) A metasurface-based PTC made from a square lattice of multilayered spheres supporting qBICs, as shown in (A). (\textbf{C}) The band structure of the metasurface from (B) under static conditions, {\textit i.e.}, $m=0$ in the spectral domain of interest. The gray-dashed lines represent the light lines, and the red-dotted line marks the spectral location of the flat band frequency $\omega_\mathrm{\Gamma}$ at the $\Gamma$-point. (\textbf{D}) The normalized scattering cross section $\sigma_\mathrm{sca}$ of the static isolated multilayered sphere used in (B) as a function of frequency. The sphere hosts a qBIC with a quality factor of $Q\approx5\times 10^4$. (\textbf{E}) The band structure of the time-varying metasurface zoomed at the very narrow frequency range centered at $\Re \{\omega\} =\omega_{\rm m}/2$. An extremely low modulation amplitude $m=7\times 10^{-8}$ is used. (\textbf{F}) The further zoomed band structure in the highlighted region of (E) illustrates a finite momentum bandgap. (\textbf{G}) The corresponding imaginary part of the frequency inside the bandgap in (F) for a fixed $\Re{(\omega)/\omega_\mathrm{m}}=0.5$. The displayed modes have a symmetry compatible with that of $p$-polarized plane waves. (\textbf{H}) The imaginary part of the frequency inside the full bandgap for a fixed $\Re{(\omega)/\omega_\mathrm{m}}=0.5$. Here, $m=3.5\times 10^{-4}$ is used. The transparency of the bands in (C), (E), and (F) is directly proportional to the radiative losses that the underlying modes possess. Moreover, $c_0$ is the speed of light in a vacuum.}
\label{fig:sphereMS}
\end{figure*}

To begin with, we analyze the BICs of static isolated multilayered spheres. We consider a sphere that consists of a core and two layers (see the inset in Fig.~\ref{fig:sphereMS}(A)). 
The core and the outer layer of the sphere are made from AZO. The dispersive permittivity of AZO is expressed using the Drude-Lorentz model (see Supplementary Section~\ref*{supp-sec:kerr}), where the corresponding parameters were taken from \cite{caspani2016enhanced}. The central layer is made from silica with a permittivity of $2.15$. We denote the radii of the core, central, and outer layers of the sphere by $r_0,\,r_1, \textrm{and } r_2$, respectively. We assume $r_0=120$~nm, $r_1=\eta r_2$, and $r_2=200$~nm, where $\eta$ is a variable. To demonstrate the existence of a true BIC, we first turn off the material losses in the AZO layers by setting the Drude-Lorentz loss parameter $\gamma$ to $0$. Next, we plot the normalized scattering cross section $\sigma_\mathrm{sca}$ of the lossless sphere as a function of the frequency $\omega$ and the geometric parameter $\eta$ (see Fig.~\ref{fig:sphereMS}(A)). The bright and dark regions in the color plot correspond to scattering zeros and resonances of the sphere, respectively. In the dashed box, the scattering zeros and the resonances approach each other, eventually merging at a point at $\eta=\eta_\mathrm{BIC}\approx0.6$. This coalescing of the zeros and the resonances is the signature for the existence of a true BIC at that point in parameter space~\cite{monticone2014embedded}. The BIC appears at the frequency where the permittivity of the AZO layers is zero \cite{monticone2014embedded}. In the following, we study how this BIC affects the band structures of metasurfaces made from the multilayered spheres. Note that there exist other geometries, such as isolated core-shell spheres, that also support BICs, which are formed due to the presence of a vanishing-permittivity layer \cite{silveirinha2014trapping}. 

The considered metasurface is made from a square lattice of multilayered spheres (see Fig.~\ref{fig:sphereMS}(B)). The lattice period of the metasurface is $\Lambda = 3r_2$. At first, we study the static case. We operate at $\eta=1.018\eta_\mathrm{BIC}=0.611$ so that the spheres support a qBIC (see Fig.~\ref{fig:sphereMS}(A)). The band structure of the corresponding static metasurface is shown in Fig.~\ref{fig:sphereMS}(C). We use the T-matrix method (based on the Mie theory) to compute the band structure (see Methods) \cite{waterman1965proceedings,garg2022modeling}. Here, we observe the emergence of flat bands due to the qBIC of the sphere. For comparison, we plot $\sigma_\mathrm{sca}$ of the sphere as a function of the frequency $\omega$ in Fig.~\ref{fig:sphereMS}(D). We note the existence of the qBIC of the sphere with the Q-factor approximately $5\times 10^4$ at the frequency $\omega_\mathrm{qBIC}$ in Fig.~\ref{fig:sphereMS}(D), which explains the appearance of the flat bands in Fig.~\ref{fig:sphereMS}(C). The shown qBIC has an electric dipolar character. Due to the spatial symmetry of the sphere, the qBIC resonance is threefold degenerate. Upon placing the sphere in the lattice, this degeneracy is lifted. Therefore, three distinct bands exist in Fig.~\ref{fig:sphereMS}(C). The flat band frequency at the $\Gamma$-point is $\omega_\mathrm{\Gamma}\approx\omega_\mathrm{qBIC}$.

Next, we turn on the time modulation of the spheres. We assume that the background permittivity $\Delta \varepsilon$ of both AZO layers of the spheres is time-periodic, \textit{i.e.}, $\Delta \varepsilon(t)=\varepsilon_\infty\left[1+m(1+\mathrm{cos}(\omega_\mathrm{m}t)\right]$ (see Supplementary Section~\ref*{supp-sec:kerr} and \ref*{supp-sec:homogeneous-pump}). The silica layer remains static. We start by considering a very small value of $m$, \textit{i.e.}, $m=7\times10^{-8}$. Note that, in addition to introducing time modulation, the pump wave offsets the background permittivity by an amount $m\varepsilon_\infty$. This offset leads to a small DC shift in the initial flat band frequency $\omega_\mathrm{\Gamma}$. The DC-shifted flat band frequency is denoted by $\omega'_\mathrm{\Gamma}$. Next, we choose the modulation frequency $\omega_\mathrm{m}$ such that $0.5\omega_\mathrm{m}=\omega'_\Gamma$. The numerical value of the modulation frequency is $\omega_\mathrm{m}=2\omega'_\Gamma=2\pi\times 459.4$~THz. The band structure of the corresponding time-varying metasurface is shown in Fig.~\ref{fig:sphereMS}(E). We observe that the band structure of the time-varying metasurface is formed by folding the band structure of the static metasurface about $\Re({\omega})=0.5\omega_\mathrm{m}$. In the time-varying case, the eigenmodes at $\Re({\omega})=0.5\omega_\mathrm{m}$ interact with each other. Choosing $0.5\omega_\mathrm{m}$ at the flat band frequency $\omega'_\mathrm{\Gamma}$ enhances this interaction due to the small radiative losses of the eigenmodes of the flat band. Therefore, light interacts with the time-varying matter for a longer duration, leading to the emergence of a noticeable momentum bandgap at $0.5\omega_\mathrm{m}$ even for vanishingly small $m$. 
We show the momentum bandgap in the magnified region around the $\Gamma$-point (Fig.~\ref{fig:sphereMS}(F)). Since the bandgap lies within the light cone, \textit{i.e.}, above the light lines, of the metasurface, we use the incident angle range $\Delta \theta$ corresponding to the $\mathbf{k}_\parallel$ values that fall inside the bandgap to quantify the gap size. In particular, the shown gap corresponds to $\Delta \theta =4^\circ$. This gap is a noticeable momentum bandgap since, even with $m=1$, gaps of comparable size do not occur in effectively homogeneous PTCs (see Supplementary Section~\ref*{supp-sec:homogenous-ptc}). Furthermore, we show the imaginary part of the frequency $\Im({\omega})$ inside the momentum bandgap for a fixed $\Re({\omega})/\omega_\mathrm{m}=0.5$ in Fig.~\ref{fig:sphereMS}(G). As expected, we observe that for the $\mathbf{k}_\parallel$ values falling inside the gap, there exist modes with $\Im({\omega})>0$ responsible for wave amplification. Here, the plot is not symmetric about $\Im{(\omega)}=0$ due to the finite (although small) radiation loss in the metasurface.

Another interesting characteristic of the qBIC-assisted PTCs is that they can support full momentum bandgaps spanning over all $\mathbf{k}_\parallel$ values. To open such a full bandgap, we increase the modulation amplitude to $m=3.5\times 10^{-4}$. The increased $m$ leads to a slightly different flat band frequency $\omega'_\mathrm{\Gamma}$. Therefore, the modulation frequency is modified to $\omega_\mathrm{m}=2\omega'_\mathrm{\Gamma}=2\pi\times 459.3$~THz. All other parameters are kept unchanged. The corresponding imaginary part of the frequency $\Im({\omega})$ inside the momentum bandgap for a fixed $\Re({\omega})/\omega_\mathrm{m}=0.5$ is shown in Fig.~\ref{fig:sphereMS}(H). We find that there exist modes with $\Im({\omega})>0$ for all possible $\mathbf{k}_\parallel$ values, leading to a full bandgap. In Fig.~\ref{fig:sphereMS}(H), we show that $\Im({\omega})>0$ on the edges of the irreducible Brillouin zone (IBZ). However, here, as well as in the following examples with full bandgaps, we verified that $\Im({\omega})>0$ inside the entire IBZ. Note that our method enhances the $\mathbf{k}_\parallel$ range for which the input waves can be amplified. However, similar to other PTC implementations, the frequency $\omega$ that gets amplified depends on the temporal boundary conditions. As discussed in \cite{khurgin2024photonic}, if the input wave (with its $\mathbf{k}_\parallel$ inside the bandgap) exists in the PTC \textit{prior} to the commencement of the time modulation, it gets amplified regardless of its frequency $\omega$. On the other hand, if the input wave enters the PTC \textit{after} the modulation has commenced, it gets amplified only if its frequency is $\omega=0.5\omega_\mathrm{m}$.

Next, we show the effects of varying the Q-factor on the required modulation amplitude to maintain the bandgap. For a fair comparison, we always require a full bandgap. First, we vary the Q-factor by changing the geometry parameter $\eta$. We note that, to open a full bandgap using the time-varying metasurface, a threshold modulation amplitude $m=m_\mathrm{th}$ is required.
\begin{figure*}
\centerline{\includegraphics[width= 1\columnwidth,trim=1 1 0.5 1,clip]{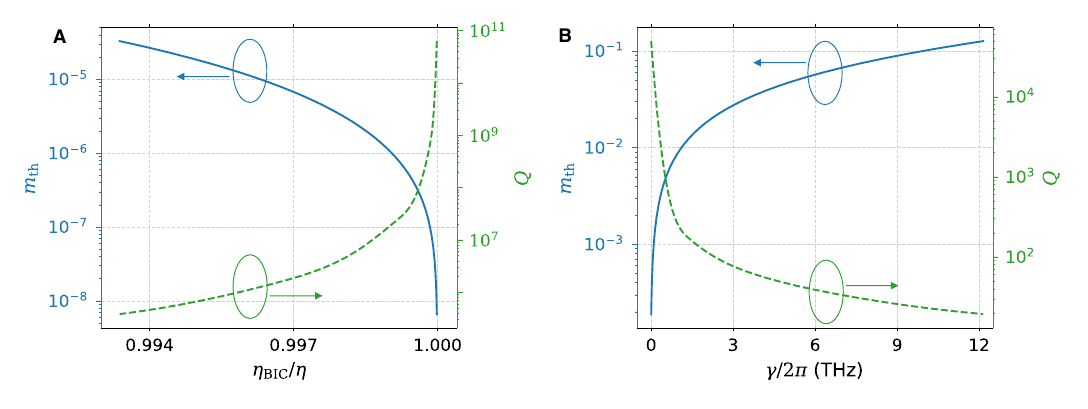}}
\caption{\textbf{Variation of the threshold modulation amplitude and the Q-factor with system parameters.} The threshold modulation amplitude $m_\mathrm{th}$ of the multilayered sphere metasurface required to exhibit a full bandgap as a function of the geometry parameter $\eta=r_1/r_2$ (blue curve in (A)) and the Drude-Lorentz loss parameter $\gamma$ (blue curve in (B)). The Q-factor of the qBIC supported by the isolated static multilayered sphere as a function of $\eta$ (green curve in (A)) and $\gamma$ (green curve in (B)). Note that $\eta/\eta_\mathrm{BIC}=1.018$ is used in (B).}
\label{fig:qvariation}
\end{figure*}
In Fig.~\ref{fig:qvariation}(A), we show $m_\mathrm{th}$ as a function of $\eta$ (blue curve). We find that as $\eta\rightarrow \eta_\mathrm{BIC}$, $m_\mathrm{th}\rightarrow 0$. This behavior is observed because, as $\eta\rightarrow \eta_\mathrm{BIC}$, the Q-factor of the qBIC approaches infinity, {\textit i.e.}, $Q\rightarrow\infty$ (green curve in Fig.~\ref{fig:qvariation}(A)). The high Q-factors reduce the radiation losses in the otherwise lossless system. Therefore, an arbitrarily small $m_\mathrm{th}$ is sufficient to maintain a full bandgap. 

So far, for a theoretical demonstration, we assumed the absence of material losses in the AZO layers. Next, we proceed to analyze the effect of material losses on $m_\mathrm{th}$ to open a full bandgap. We plot $m_\mathrm{th}$ as a function of the Drude-Lorentz loss parameter $\gamma$ of the AZO layers in Fig.~\ref{fig:qvariation}(B) (blue curve). Note that $\eta/\eta_\mathrm{BIC}$ is fixed as $\eta/\eta_\mathrm{BIC}=1.018$. We observe that upon increasing $\gamma$, the required $m_\mathrm{th}$ increases. This behavior is expected because increasing the material losses decreases the Q-factor of the qBICs (green curve in Fig.~\ref{fig:qvariation}(B)). Therefore, a higher $m_\mathrm{th}$ is needed to open a full bandgap.

In particular, the experimentally reported value of $\gamma$ in the spectral region of interest (the ENZ regime of AZO) is $\gamma\approx2\pi\times12$~THz \cite{caspani2016enhanced}. From Fig.~\ref{fig:qvariation}(B), we note that $m_\mathrm{th}=0.13$ is required to open a full bandgap using this value of $\gamma$. The $m_\mathrm{th}$ value corresponds to a pump intensity of $1.3$~$\mathrm{TW/cm^2}$. In comparison, effectively homogeneous PTCs require $m_\mathrm{th}\sim 1$ to even open finite bandgaps (see Supplementary Section~\ref*{supp-sec:homogenous-ptc}).

It is worth mentioning that, while the qBICs of the multilayered spheres assist in decreasing $m_\mathrm{th}$, assuming realistic losses, the corresponding metasurfaces still require rather high modulation amplitudes (and pump intensities) to open full bandgaps. In literature, metamaterial-based approaches have shown that the losses in the effective ENZ medium can be considerably reduced \cite{Kelly2019engineering}. Such an approach could be beneficial while using multilayered sphere qBICs to attain wide momentum bandgaps while maintaining low modulation amplitudes. Another challenge with multilayered sphere geometry is that it is difficult to fabricate. 
Keeping these difficulties in mind, we present a more practical design of the PTCs that utilizes the qBICs of dielectric cylinders as follows.

\subsection*{PTCs based on metasurfaces made from germanium cylinders}
Now, we begin with an isolated static germanium (Ge) cylinder supporting a high-Q qBIC. The material dispersion in the Ge cylinder is considered using the Drude-Lorentz model (see Supplementary Section~\ref*{supp-sec:kerr}). The Drude-Lorentz parameters of Ge are taken from \cite{Amotchkina2020characterization}. Realistic material losses in the Ge cylinder are included from the outset. The cylinder, with a radius of $r = 1030$~nm and an aspect ratio of $r/h \approx 0.703$ (where $h$ is the cylinder height), has been geometrically optimized following the strategy discussed in \cite{rybin2017high,koshelev2020subwavelength}. This optimization leverages an avoided crossing between a Mie-type mode of azimuthal index $m=0$ and principal (radial) index $n=1$ and a neighboring $m = 0$ Mie-type mode of different principal order ($n=2$), where their interaction suppresses the radiative losses in the resulting hybrid mode. 

Group theory analysis reveals that these $m = 0$ modes belong to the $A_{2g}$ irreducible representation of the $D_{\infty h}$ point group that is relevant here. As a result, the hybrid mode is symmetry-protected and dark at axial incidence but emerges as a sharp qBIC under non-axial $s$-polarized plane wave excitation \cite{Gantzounis2009plasmon}. The qBIC is formed by the Friedrich-Wintgen mechanism \cite{rybin2017high}. We highlight that this qBIC remains robust against the thermal expansion and thermo-optic effects for the considered pump intensities in the following (see Supplementary Section~\ref*{supp-sec:static-cylinder-bic}). Furthermore, because the qBIC has a symmetry compatible only with that of $s$-polarized plane waves, no resonance peak is observed for the $p$-polarized plane wave excitation. 

\begin{figure*}
\centerline{\includegraphics[width= 1\columnwidth,trim=0.1 0.1 0.1 1,clip]{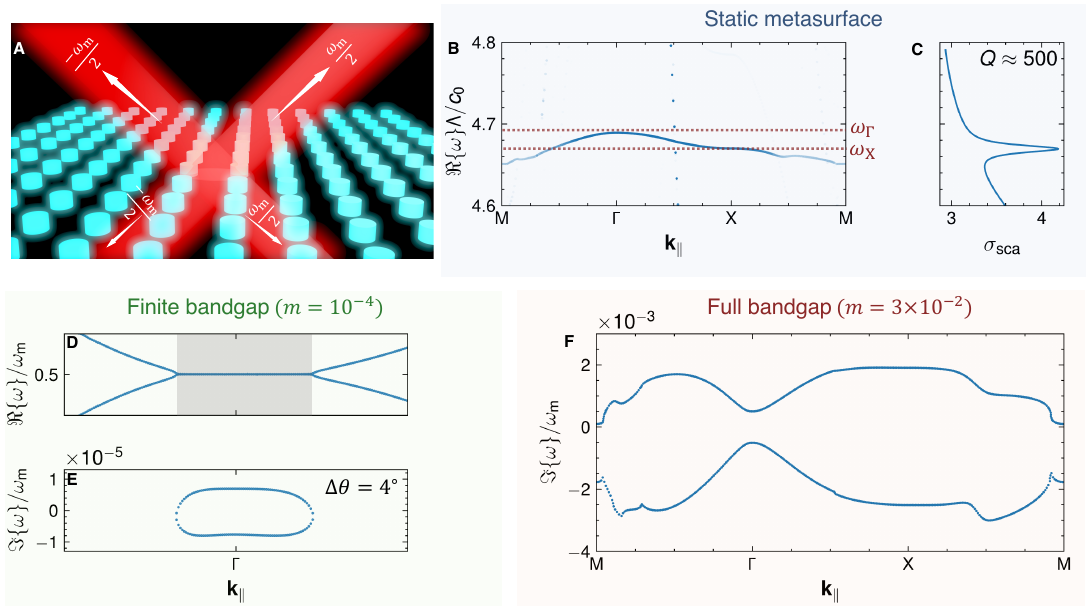}}
\caption{\textbf{PTC based on cylinder metasurface.} (\textbf{A}) A metasurface-based PTC made from a square lattice of resonant cylinders. (\textbf{B}) The band structure of the metasurface shown in (A) under static conditions, {\textit i.e.}, $m=0$. Here, all the shown modes lie above the light lines. The red-dotted lines mark the flat band frequencies $\omega_\Gamma$ and $\omega_\mathrm{X}$ at the $\Gamma$ and X-points, respectively. (\textbf{C}) The rotationally averaged normalized scattering cross section $\sigma_\mathrm{sca}$ of the isolated cylinder used in (A) as a function of frequency. The cylinder hosts a qBIC with the quality factor $Q\approx500$. (\textbf{D}) The band structure of the time-varying metasurface zoomed at the very narrow frequency range centered at $\Re \{\omega\} =\omega_{\rm m}/2$ near the $\Gamma$-point, illustrating a finite bandgap. Here, $m=10^{-4}$ is used. (\textbf{E}) The corresponding imaginary part of the frequency inside the bandgap in (D) for a fixed $\Re{(\omega)/\omega_\mathrm{m}}=0.5$. The displayed mode branch has a symmetry compatible with that of $s$-polarized plane waves. (\textbf{F}) The imaginary part of the frequency inside the full bandgap for a fixed $\Re{(\omega)/\omega_\mathrm{m}}=0.5$. Here, $m=3\times10^{-2}$ is used.}
\label{fig:cylinderqbic}
\end{figure*}

First, we analyze static metasurfaces made from such qBIC-supporting cylinders (see Fig.~\ref{fig:cylinderqbic}(A)). We assume the lattice period of the metasurface to be $\Lambda=3.2r$. The band structure of the metasurface under static conditions is shown in Fig.~\ref{fig:cylinderqbic}(B), where we observe the emergence of a flat band. The flat band frequencies at the $\Gamma$ and X-points are $\omega_{\Gamma}$ and $\omega_\mathrm{X}$, respectively. This flat band arises due to the existence of the aforementioned qBIC that the isolated static Ge cylinder hosts. For comparison, we show the rotationally averaged normalized scattering cross section $\sigma_\mathrm{sca}$ of the cylinder as a function of frequency $\omega$ in Fig.~\ref{fig:cylinderqbic}(C). In Fig.~\ref{fig:cylinderqbic}(C), we observe the existence of the qBIC with $Q\approx500$, which explains the existence of the flat band in Fig.~\ref{fig:cylinderqbic}(B). Note that the Q-factor of the qBIC of the Ge cylinders is considerably lower than that of the multilayered spheres. This is because the spheres were assumed to be lossless. With realistic material losses, as here, finite dielectric nanoresonators cannot achieve arbitrarily high Q-factors \cite{koshelev2020subwavelength}. The Q-factor of the qBIC of a dielectric nanoresonator goes to infinity in the limit of vanishing material losses and for a diverging real part of the permittivity \cite{rybin2017high}. However, as we demonstrate below, the Q-factor of the considered cylinders with realistic material parameters is sufficiently high to substantially reduce the required modulation amplitude for opening noticeable momentum bandgaps, compared to effectively homogeneous PTCs and dipolar Mie-resonant metasurfaces. 

Next, we turn on the time modulation of the metasurface. We assume the background permittivity $\Delta\varepsilon$ of the Ge cylinders to be time-periodic. The modulation amplitude is assumed to be $m=10^{-4}$. This value of $m$ corresponds to a pump intensity of $0.7$~$\mathrm{GW/cm^2}$. First, we aim at opening a finite bandgap near the $\Gamma$-point. Therefore, we choose the modulation frequency $\omega_\mathrm{m}$ such that $0.5\omega_\mathrm{m}=\omega'_{\Gamma}$. Here, $\omega'_{\Gamma}$ is the DC-shifted flat band frequency at the $\Gamma$-point. Numerically, $\omega_\mathrm{m}=2\omega'_{\Gamma}=2\pi\times 136$~THz.
We show the noticeable momentum bandgap hosted by the metasurface in the region around the $\Gamma$-point in Fig.~\ref{fig:cylinderqbic}(D). We also plot the imaginary part of the frequency inside the bandgap in Fig.~\ref{fig:cylinderqbic}(E). The displayed bandgap corresponds to an incident angle range of $\Delta\theta=4^\circ$. We find that the cylinder-based metasurface requires an order of magnitude lower modulation amplitude than the recently reported dipolar Mie resonant structure to achieve a bandgap of the same size (see Supplementary Section~\ref*{supp-sec:MD-sph}) \cite{wang2025expanding}. Moreover, even with $m=1$, such a gap size is not attainable in effectively homogeneous PTCs (see Supplementary Section~\ref*{supp-sec:homogenous-ptc}). Importantly, the laser-induced damage threshold intensity of bulk Ge is $94$~$\mathrm{GW/cm^2}$ using multi-shot pump pulses \cite{malik2013determination}. Furthermore, the nanostructuring of Ge reduces its damage threshold by an order of magnitude, \textit {i.e.},to a pump intensity of $9.4$~$\mathrm{GW/cm^2}$ \cite{grinblat2017efficient,kern2011comparison}. Therefore, the modulation amplitude of $m=10^{-4}$ corresponding to the pump intensity of $0.7$~$\mathrm{GW/cm^2}$ is well reachable in Ge using the optical Kerr effect without causing laser-induced damage to the underlying material. Note that while Ge is almost transparent at $0.5\omega_\mathrm{m}$, time modulation-induced upconverted frequencies may lie in its high material loss regime. Our T-matrix-based formalism accounts for such losses in all the examples studied in this work. In fact, at low modulation amplitudes, the coupling to those higher frequencies remains rather weak. Therefore, those material losses do not impact $\Im(\omega)$ inside the bandgap appreciably.

Finally, we show that, similar to the previous example, time-varying metasurfaces based on cylinders can also support full bandgaps. To illustrate this, we increase the modulation amplitude of the metasurface to $m=3\times10^{-2}$. We choose the modulation frequency $\omega_\mathrm{m}$ such that $0.5\omega_\mathrm{m}=\omega'_\mathrm{X}$. Here, $\omega'_\mathrm{X}$ is the DC-shifted frequency of the eigenmode at the X-point. Our simulations show that this choice of $\omega_{\mathrm{m}}$ requires the smallest modulation amplitude $m$ to open a full bandgap, as it maximizes the average overlap between the original modes and the band-folded modes induced by time modulation. Numerically, $\omega_\mathrm{m}=2\omega'_\mathrm{X}=2\pi\times 134$~THz. All other parameters are unchanged. We plot the imaginary part of the frequency inside the bandgap in Fig.~\ref{fig:cylinderqbic}(F). We observe the existence of modes with $\Im{(\omega)>0}$ for all $\mathbf{k}_\parallel$, leading to the demonstration of a full bandgap. The chosen value of $m$ corresponds to a pump intensity of $210$~$\mathrm{GW/cm^2}$. This pump intensity is about $20$ times higher than the damage threshold of structured Ge under multi-shot laser pulses. However, using pump pulses with a reduced repetition rate can considerably increase the damage threshold \cite{Austin2018femtosecond}. Therefore, a careful choice of the repetition rate may bring full bandgaps in reach in future works.

\section*{Discussion}
In conclusion, we have shown that by capitalizing on qBICs sustained in meta-atoms, we can substantially reduce the required modulation amplitude to observe noticeable momentum bandgaps in the corresponding metasurface-based PTCs. Using two different structures, we have shown the impact of \textit{material-induced} as well as \textit{geometry-induced} qBICs. Our theoretical analysis, based on metasurfaces composed of time-varying multilayered spheres, demonstrates that, by suppressing radiative losses in the meta-atoms, it is possible to achieve large momentum bandgaps for arbitrarily small modulation amplitudes, provided that absorption losses are negligible. Furthermore, we explored a viable route to realize a PTC based on metasurfaces that exploit qBICs supported by Ge cylinders. Using these metasurfaces, we demonstrate noticeable momentum bandgaps for modulation amplitudes as small as $m=10^{-4}$ corresponding to a pump intensity of $0.7$~$\mathrm{GW/cm^2}$. This modulation amplitude is many orders of magnitude lower than that required in conventional homogeneous PTCs to open similarly-sized bandgaps. This value of $m$ is achievable using the optical Kerr effect without causing laser-induced damage to the Ge cylinders. We anticipate that our results will be helpful in the experimental realization of PTCs at optical frequencies. We emphasize here that, apart from the discussed structures, the proposed framework can also be applied to other high-Q geometries to attain comparable results. These include structures with higher-order Mie resonances and those with more complicated meta-atom designs. In general, the threshold modulation amplitude $m_\mathrm{th}$ required to open noticeable bandgaps decreases with increasing Q-factor. Moreover, we presented the application of the Mie theory to the time-varying metasurfaces made from multilayered spheres and finite-sized cylinders. We believe that our proposed numerical scheme will serve as a valuable tool for the further exploration of structured PTCs.
\section*{Methods}

\subsection*{Band structure calculation}
We employ the T-matrix method to evaluate the band structures of the time-varying metasurfaces \cite{garg2022modeling,panagiotidis2023optical}. The T-matrix is closely linked to the well-known S-matrix used in scattering theory (See Supplementary Section~\ref*{supp-sec:smat})\cite{waterman1976matrix}. According to the method, the incident field coefficients $\mathbf{A}_\mathrm{inc}$ and the scattered field coefficients $\mathbf{A}_\mathrm{sca}$, computed in a basis of vector spherical harmonics (VSHs), satisfy
\begin{equation}
\mathbf{A}_\mathrm{sca}= \left(\mathbf{\hat{I}}-{\mathbf{\hat{T}}_{0}(\omega)}\sum_{\mathbf{R}\neq0}\mathbf{\hat{C}}^{(3)}(-\mathbf{R})\hspace{1pt}\mathrm{e}^{i\mathbf{k_{{\parallel}}}\cdot\mathbf{R}}\right)^{-1}{\mathbf{\hat{T}}_{0}(\omega)}\mathbf{A}_\mathrm{inc}\,.\label{eq:Teff}
\end{equation}
Here, $\mathbf{\hat{I}}$ is the identity matrix, $\mathbf{R}$ is a lattice vector, and $\mathbf{\hat{C}}^{(3)}$ is the translation matrix of VSHs \cite{beutel2021efficient, beutel2024treams,garg2022modeling}. Furthermore, ${\mathbf{\hat{T}}}_0$ is the T-matrix of an isolated time-varying scatterer. The T-matrices of isolated time-varying multilayered spheres can be evaluated using the Floquet-Mie theory, while those of isolated time-varying cylinders can be computed using the dynamic extended boundary condition method (EBCM) \cite{lamprianidis2023generalized,ptitcyn2023floquet,Stefanou2023light}. Moreover, the lattice summation shown in Eq.~\eqref{eq:Teff} is calculated using the \texttt{treams} package \cite{beutel2024treams}. To allow evaluating the lattice sums in the lower half of the complex frequency plane ({\textit i.e.}, for $\Im({\omega})<0$), the branch cut of the incomplete gamma function is rotated down from the typical negative real axis by $\pi/2$. The incomplete gamma function with shifted branch cut is evaluated via its relation to the unshifted incomplete gamma function (see Eq. (8.2.9) of reference \cite{NIST:DLMF}). Here, we use $J$ as the total number of frequencies that the time-varying structure scatters upon a monochromatic incidence to construct $\mathbf{\hat{C}}^{(3)}$ and ${\mathbf{\hat{T}}}_0$. Furthermore, these matrices consist of terms up to the maximum multipolar order $l_\mathrm{max}$. In particular, for computing the T-matrix of cylinders, using the dynamic EBCM, we use the maximum multipolar order $l_\mathrm{cut}$. Here, $l_\mathrm{cut}>l_\mathrm{max}$ is chosen to ensure sufficient numerical convergence. However, after computing the T-matrix, we retain its terms only up to the order $l_\mathrm{max}$. Therefore, the dimension of the matrices $\mathbf{\hat{C}}^{(3)}$ and ${\mathbf{\hat{T}}}_0$ is $2J\hspace{1pt}l_\mathrm{max}(l_\mathrm{max}+2)$. We provide the used $J$, $l_\mathrm{max}$, and $l_\mathrm{cut}$ values for generating the results in this article in the Supplementary Section~\ref*{supp-sec:convergence-params}.

For calculating the spectral location of an eigenmode, we require $\mathbf{A}_\mathrm{sca}\neq0$ while $\mathbf{A}_\mathrm{inc}=0$ \cite{rahimzadegan2022comprehensive, wolff2018modal}. Using Eq.~\eqref{eq:Teff}, we find that this requirement is satisfied if the following holds

\begin{equation}
\label{eq: Tmat_BS}
\underbrace{
\left|\mathbf{\hat{I}}-\mathbf{\hat{T}}_0(\omega)\sum_{\mathbf{R}\neq0}\mathbf{\hat{C}}^{(3)}(-\mathbf{R})\hspace{1pt}\mathrm{e}^{i\mathbf{k_{\parallel}}\cdot\mathbf{R}}\right|}_{\left| \mathbf{\Psi}\left(\mathbf{k_{\parallel}}, \omega \right)\right|}=0\,.
\end{equation}
\noindent
Therefore, those $\omega$ and $\mathbf{k_{\parallel}}$ values that satisfy Eq.~\eqref{eq: Tmat_BS} characterize the spectral location of an eigenmode in the band structure of the metasurface. The band structure is plotted as a scatter plot of the spectral locations of the eigenmodes of the metasurface. Due to radiative loss and the gain/loss introduced by the time modulation, the bands generally possess complex-valued $\omega$. Locating the resonance frequencies of eigenmodes is a well-known computational problem for static photonic resonators. Recently, the adaptive Antoulas--Anderson (AAA) algorithm \cite{nakatsukasa_aaa_2018} has been introduced to the field of resonant nanophotonics \cite{betz_efficient_2024} to locate the eigenfrequencies as poles of some optical response function. The AAA algorithm was shown to be well-suited for locating the resonances of coupled clusters with a finite number of meta-atoms, leveraging the T-matrix method \cite{fischbach_framework_2025}. Here, we generalize this approach to periodic lattices of time-varying meta-atoms. The AAA algorithm is used to find a good rational approximation $r_\mathbf{k_{\parallel}}(\omega)$ given by
$$r_\mathbf{k_{\parallel}}(\omega) \approx f\left(\mathbf{k_{\parallel}}, \omega\right) = \frac{1}{\left| \mathbf{\Psi}\left(\mathbf{k_{\parallel}}, \omega \right)\right|}\, ,$$
based on a sufficient number of sample evaluations of $\left| \mathbf{\Psi}\left(\mathbf{k_{\parallel}}, \omega \right)\right|$ for a fixed $\mathbf{k_{\parallel}}$. Then, performing a pole search of the function $r_\mathbf{k_{\parallel}}(\omega)$ allows us to find the $\omega$ and $\mathbf{k}_\parallel$ values that satisfy Eq.~\eqref{eq: Tmat_BS}.


\clearpage 

%
\bibliography{usedrefs} 
\bibliographystyle{sciencemag}
%
%
%
%
%
%


\paragraph*{Acknowledgments:}
The authors would like to thank Dr. Kirill Koshelev for the fruitful discussions about qBICs.  
\paragraph*{Funding:}
V.A. acknowledges the Finnish Foundation for Technology Promotion, and Research Council of Finland Flagship Programme, Photonics Research and Innovation (PREIN), decision number 346529, Aalto University.
X.W. acknowledges the Fundamental Research Funds for the Central Universities, China (project no. 3072024WD2603).
P.G. and C.R. are part of the Max Planck School of Photonics, supported by the Bundesministerium für Bildung und Forschung, the Max Planck Society, and the Fraunhofer Society. P.G. and C.R. acknowledge support by the German Research Foundation within the SFB 1173 (project ID no. 258734477). P.G. and J.D.F. acknowledge support from the Karlsruhe School of Optics and Photonics (KSOP). M.N. and C.R. acknowledge support by the KIT through the “Virtual Materials Design” (VIRTMAT) project.
J.D.F. and C.R. acknowledge financial support by the Helmholtz Association in the framework of the innovation platform “Solar TAP”.
\paragraph*{Author contributions:\\}
Conceptualization: P.G., C.R., V.A.\\
Methodology: P.G., C.R., M.N., N.S., X.W., J.D.F., M.S.M. \\
Software: P.G., L.Z., M.N., E.A., N.S., N.P., J.D.F. \\
Validation: P.G., E.A., N.S., N.P.\\
Formal Analysis: P.G.\\
Investigation: P.G., E.A., N.S., N.P., J.D.F.\\
Visualization: P.G., X.W.\\
Supervision: C.R., V.A. \\
Project administration: C.R.\\
Writing—original draft: P.G., C.R., V.A. \\
Writing—review \& editing: P.G., C.R., M.N., E.A., N.S., N.P., V.A., X.W, J.D.F., M.S.M.\\
Funding Acquisition: C.R. 
\paragraph*{Competing interests:}
All other authors declare they have no competing
interests.
\paragraph*{Data, code, and materials availability:}
All data and code needed to evaluate and reproduce the results in the paper are present in the paper and the Supplementary Materials. This study did not generate new materials.


\subsection*{Supplementary materials}
Supplementary Text\\
Figures S1 to S6\\


\newpage


\renewcommand{\thefigure}{S\arabic{figure}}
\renewcommand{\thetable}{S\arabic{table}}
\renewcommand{\theequation}{S\arabic{equation}}
\renewcommand{\thepage}{S\arabic{page}}
\setcounter{figure}{0}
\setcounter{table}{0}
\setcounter{equation}{0}
\setcounter{page}{1} 


\begin{center}
\section*{Supplementary Materials for the paper\\ ``\scititle"}

	Puneet Garg$^{1\ast}$,
	Evangelos Almpanis$^{2,3}$,
	Leander Zimmer$^{1}$,
    Jan David Fischbach$^{4}$, \and
    Xuchen Wang$^{5}$,
    Mohammad S. Mirmoosa$^{6}$,
    Markus Nyman$^{1}$,
    Nikolaos Stefanou$^{2}$, \and
    Nikolaos Papanikolaou$^{3}$,\\
    Viktar Asadchy$^{7\ast}$,
    Carsten Rockstuhl$^{1,4\ast}$\\
	\small$^\ast$Corresponding authors. Email: puneet.garg@kit.edu, viktar.asadchy@aalto.fi, carsten.rockstuhl@kit.edu 
\end{center}

\subsubsection*{This PDF file includes:}
Supplementary Text\\
Figures S1 to S6

\newpage 
\clearpage 
\section*{Supplementary Text}
\setcounter{section}{0}
\renewcommand{\thesection}{S\arabic{section}}
\setcounter{figure}{0}
\renewcommand{\thefigure}{S\arabic{figure}}
\newcounter{SIfig}
\renewcommand{\theSIfig}{S\arabic{SIfig}}
\setcounter{equation}{0}
\renewcommand{\theequation}{S\arabic{equation}}

\section{Time modulation due to the optical Kerr effect}\label{supp-sec:kerr}
We use the optical Kerr effect as the source of time modulation in the discussed metasurface scenarios. In a homogeneous nonlinear Kerr medium, the physical (real-valued) polarization density for the signal field can be written as \cite{khurgin2024photonic,Khurgin2021fast,xiao2014propagation}

\begin{equation}
\mathbf{P}(\mathbf{r},t)=\varepsilon_0\int_{-\infty}^{\infty} R^\mathrm{(1)}(t-\tau)\,\mathbf{E}(\mathbf{r},\tau)\mathrm{d}\tau+\varepsilon_0\chi^{(3)}E_\mathrm{pump}^2(\mathbf{r},t)\,\mathbf{E}(\mathbf{r},t)\,,\label{eq:kerrpol}    
\end{equation}
\noindent
where $\mathbf{E}$ and $E_\mathrm{pump}$ are the real-valued signal and pump electric fields, respectively. We use the undepleted pump approximation. Moreover, the polarization of the pump field is also fixed. Here, $R^{(1)}(t-\tau)$ is the dispersive linear response function, and $\chi^{(3)}$ is the third-order nonlinear susceptibility of the medium. Note that the third-order nonlinear response function of the medium is defined as $R^{(3)}(t-\tau)=\delta(t-\tau)\chi^{(3)}$, where the Dirac delta distribution, $\delta(t-\tau)$, represents the instantaneous nature of the third-order nonlinear response of the medium to the electric fields \cite{Khurgin2021fast}. Physically, this instantaneous response is facilitated by the virtual excitation and relaxation of the electrons in the nonlinear medium \cite{Khurgin2021fast}. Note that the polarization density model in Eq.~\eqref{eq:kerrpol} to describe the physics of PTCs has been proposed in \cite{khurgin2024photonic,Khurgin2021fast,xiao2014propagation}. Next, we use a pump field given by $E_\mathrm{pump}=E_\mathrm{p}(\mathbf{r})\mathrm{cos}(\omega_\mathrm{p}t+\phi(\mathbf{r}))$. Therefore, Eq.~\eqref{eq:kerrpol} can be written as

\begin{align}
\mathbf{P}(\mathbf{r},t)&=\varepsilon_0\int_{-\infty}^{\infty} R^\mathrm{(1)}(t-\tau)\,\mathbf{E}(\mathbf{r},\tau)\mathrm{d}\tau+\varepsilon_0\chi^{(3)}E_\mathrm{p}^2(\mathbf{r})\mathrm{cos}^2(\omega_\mathrm{p}t+\phi(\mathbf{r}))\,\mathbf{E}(\mathbf{r},t)\,\\
&=\varepsilon_0\int_{-\infty}^{\infty} R^\mathrm{(1)}(t-\tau)\,\mathbf{E}(\mathbf{r},\tau)\mathrm{d}\tau+\varepsilon_0\frac{\chi^{(3)}}{2}E_\mathrm{p}^2(\mathbf{r})\left[ 1+\mathrm{cos}(2\omega_\mathrm{p}t+2\phi(\mathbf{r}))\right] \,\mathbf{E}(\mathbf{r},t)\,.
\end{align}
\noindent
For simplicity, we assume that the pump field is independent of the spatial coordinate $\mathbf{r}$. We provide a justification for this homogeneous pump approximation in Supplementary Section~\ref{supp-sec:homogeneous-pump}. Therefore, we write

\begin{align}
\mathbf{P}(\mathbf{r},t)
&=\varepsilon_0\int_{-\infty}^{\infty} \left\{ R^\mathrm{(1)}(t-\tau)\,+\delta(t-\tau)\frac{\chi^{(3)}}{2}E_\mathrm{p}^2 \left[ 1+\mathrm{cos}(2\omega_\mathrm{p}t+2\phi) \right] \right\}\,\mathbf{E}(\mathbf{r},\tau)\mathrm{d}\tau\,.
\end{align}

\noindent
The optical Kerr effect gives rise to the following time-dependent effective linear response function
\begin{equation}
\label{eq:Reff_time}
R_\mathrm{eff}^{(1)}(t,t-\tau)
=R^{(1)}(t-\tau)\,+\delta(t-\tau)\frac{\chi^{(3)}}{2}E_\mathrm{p}^2 \left[1+\mathrm{cos}(2\omega_\mathrm{p}t+2\phi) \right]\,.
\end{equation}
\noindent
We rewrite $R^{(1)}(t-\tau)=\chi_\infty\delta(t-\tau)+R^{(1)}_\mathrm{disp}(t-\tau)$, where $R^{(1)}_\mathrm{disp}$ is the dispersive part of the linear response function and $\chi_\infty$ is the non-dispersive background susceptibility of the underlying static medium. Therefore

\begin{equation}\label{eq:total-response_func}
R_\mathrm{eff}^{(1)}(t,t-\tau)
=R^{(1)}_\mathrm{disp}(t-\tau)\,+\delta(t-\tau)\Delta\chi(t)\,,
\end{equation}
where
\begin{equation}
\Delta\chi(t)
=\chi_\infty+\frac{\chi^{(3)}E_\mathrm{p}^2}{2} \left[1+\mathrm{cos}(2\omega_\mathrm{p}t+2\phi) \right]\,.
\end{equation}
\noindent
Moreover, we can write 

\begin{equation}\label{eq:bg_permittivity}
\Delta\varepsilon(t)=1+\Delta\chi(t)
=\varepsilon_\infty\left\{ 1+\frac{\chi^{(3)}E_\mathrm{p}^2}{2\varepsilon_\infty}\left[1+\mathrm{cos}(2\omega_\mathrm{p}t+2\phi) \right] \right\}\,.
\end{equation}
\noindent
Here, $\Delta\chi(t)$ and $\Delta\varepsilon(t)$ denote the background susceptibility and permittivity of the time-varying medium, respectively. In addition, $\chi_\infty$ and $\varepsilon_\infty$ $(= 1 + \chi_\infty)$ represent the corresponding parameters of the underlying static medium. Note that this time-varying background permittivity model has been employed to explain the experimental results in \cite{galiffi2026optical}. Next, we define the modulation frequency $\omega_\mathrm{m}=2\omega_\mathrm{p}$, and modulation amplitude $m=\frac{\chi^{(3)}E_\mathrm{p}^2}{2\varepsilon_\infty}$. Hence
\begin{equation}
\Delta\varepsilon(t)
=\varepsilon_\infty\left\{1+m\big[1+\mathrm{cos}(\omega_\mathrm{m}t)\big]\right\}\,.
\label{eq:bg_permittivity2}
\end{equation}
\noindent
Here, the initial phase $\phi$ of the pump field is set to zero without loss of generality, as any nonzero phase merely amounts to the time-translation of $\Delta\varepsilon(t)$ by $\Delta t=2\phi/\omega_\mathrm{m}$. Such a time-translation does not affect the band structures of the PTCs. This phase insensitivity of PTCs has been discussed in detail in \cite{wang2023metasurface}. 

Next, we perform the Fourier transform of Eq.~\eqref{eq:Reff_time} to obtain \cite{ptitcyn2023floquet}

\begin{equation}\label{eq:response_func_freq_domain}
\begin{aligned}
\tilde{R}_\mathrm{eff}^{(1)}(\omega-\omega',\omega')
&= \delta(\omega-\omega')\sum_j
\frac{\omega^2_{\mathrm{p}j}}{\omega^2_{0j}-i\gamma_j-\omega^2} \\
&\quad + \chi_\infty\left\{
\delta(\omega-\omega')
+ \frac{\chi^{(3)}E_\mathrm{p}^2}{2\chi_\infty}\left[
\delta(\omega-\omega')
+ \frac{1}{2}\Big(\delta(\omega-\omega'-\omega_\mathrm{m})
+ \delta(\omega-\omega'+\omega_\mathrm{m})
\Big)\right]
\right\}.
\end{aligned}
\end{equation}
Here, $\tilde{R}_\mathrm{eff}^{(1)}(\omega-\omega',\omega')$ is the Fourier transform of the linear response function $R_\mathrm{eff}^{(1)}(t, t-\tau)$. Moreover, we use the Drude-Lorentz model to consider material dispersion, with $\omega_{\mathrm{p}j}$ being the plasma frequency, $\omega_{\mathrm{0}j}$ the resonance frequency, and $\gamma_j$ the loss parameter of the $j^\mathrm{th}$ Drude-Lorentz oscillator. Note that in the main text, we use the nomenclature $\gamma_1=\gamma$. We have used the Drude-Lorentz parameters of Ge from \cite{Amotchkina2020characterization} and $\chi^\mathrm{(3)}$ from ~\cite{Hon2011the}. For AZO, the corresponding parameters are taken from \cite{caspani2016enhanced}.

\section{Homogeneous pump approximation}\label{supp-sec:homogeneous-pump}
\begin{figure*}[h]
\centerline{\includegraphics[width= 0.6\columnwidth]{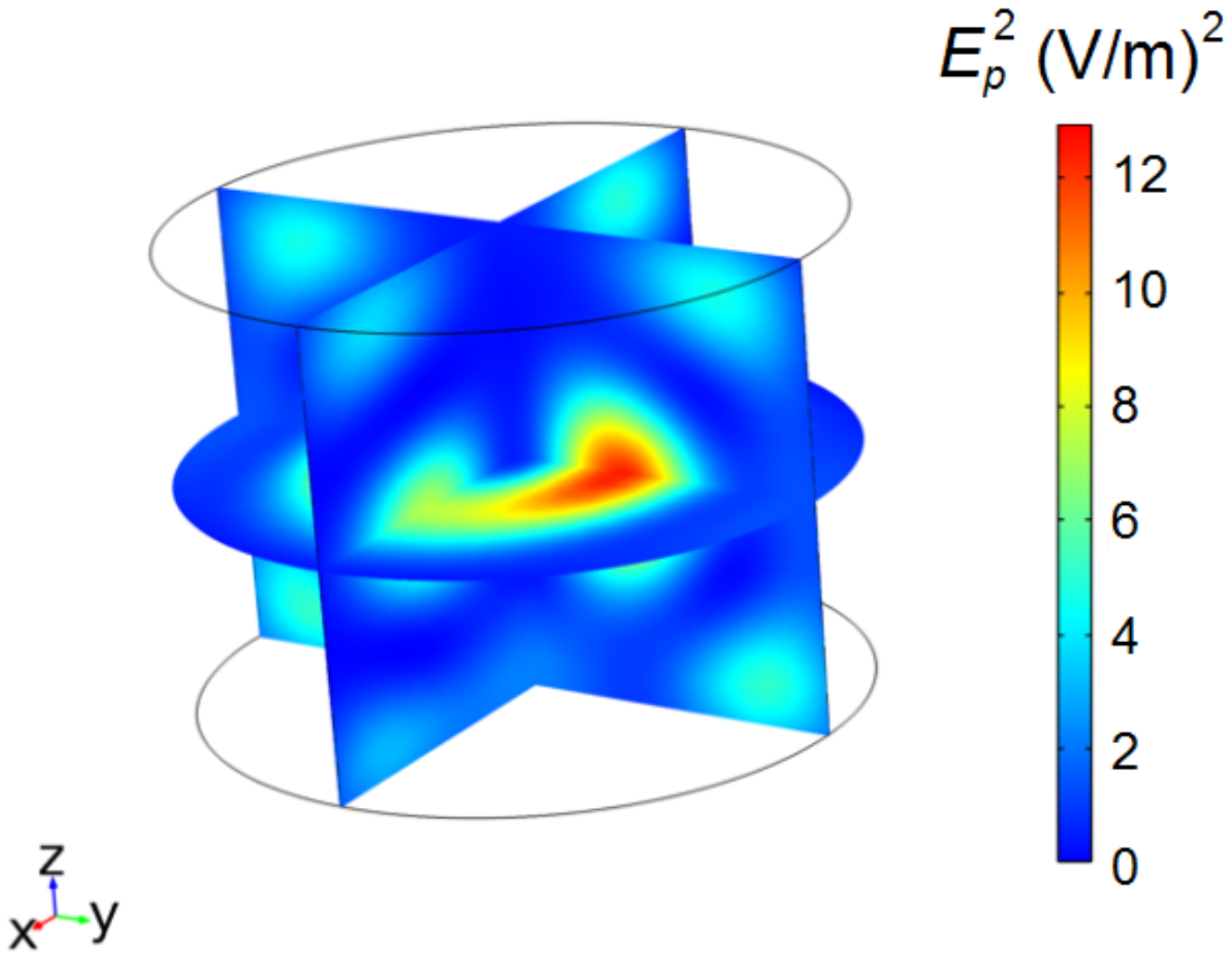}}
\caption{\textbf{Pump-induced spatial inhomogeneity of modulation amplitude}. Spatial distribution of the pump-induced electric field (squared amplitude), \textit{i.e.}, $E^2_\mathrm{p}$ inside a single cylindrical germanium nanoresonator of the metasurface at the pump frequency $\omega_\mathrm{p}=0.5\omega_\mathrm{m}$. Here, $0.5\omega_\mathrm{m}$ is chosen to be the same as that in Figs.~\ref{fig:cylinderqbic}(D)--(E). The pump field is a $p$-polarized plane wave of amplitude $1$~V/m, incident at $5^{\circ}$
with respect to the metasurface normal, a configuration that does not excite the high-Q modes of the flat band. Although the internal field exhibits spatial inhomogeneity, the Kerr-induced background permittivity modulation remains perturbative and can be effectively represented by a homogeneous pump amplitude in the full-wave calculations.}
\refstepcounter{SIfig}\label{fig:S1}
\end{figure*}
\noindent{The Kerr-induced temporal modulation experienced by the signal wave is governed by the pump field distribution inside the nanoresonators. In general, the internal pump field, and thus the induced background permittivity change, are spatially inhomogeneous. However, since the Kerr-induced background permittivity modulation remains perturbative with respect to the static background permittivity, its detailed spatial structure influences the signal dynamics only through overlap integrals between the space-dependent modulation profile $m(\mathbf{r})$ and the relevant eigenmodes of the underlying \textit{static} scattering structure. In particular, the coupling coefficient of the original mode and the time modulation-induced mode on the flat band can be written as \cite{asadchy2022parametric}

\begin{equation}
 \label{eq:kappa}
\kappa=\int_{V}\mathbf{E}^*_\mathrm{res}(\mathbf{r})\,m(\mathbf{r})\,\mathbf{E}_\mathrm{res}(\mathbf{r})\mathrm{d}V\,,
\end{equation}
where $\mathbf{E}_\mathrm{res}(\mathbf{r})$ and $\mathbf{E}^*_\mathrm{res}(\mathbf{r})$ are the spatial field profiles of the eigenmodes on the flat bands corresponding to the frequencies $0.5\omega_\mathrm{m}$ and $-0.5\omega_\mathrm{m}$, respectively, and $V$ is the volume of the nanoresonator inside the lattice. Note that $\mathbf{E}_\mathrm{res}(\mathbf{r})$ and $\mathbf{E}^*_\mathrm{res}(\mathbf{r})$ correspond to the field profiles of the high-Q modes of the underlying \textit{static} metasurface. They can be excited by the signal wave. Moreover, the pump wave couples these modes through a time modulation characterized by the space-dependent modulation amplitude profile $m(\mathbf{r})$. The formulation defined by Eq.~\eqref{eq:kappa} has been derived and numerically verified in \cite{asadchy2022parametric}. Next, Eq.~\eqref{eq:kappa} can be approximated to define an effective modulation amplitude as

\begin{equation} \label{eq:kappa2}
\kappa=\int_{V}\mathbf{E}^*_\mathrm{res}(\mathbf{r})\,m(\mathbf{r})\,\mathbf{E}_\mathrm{res}(\mathbf{r})\mathrm{d}V\approx\int_{V}\mathbf{E}^*_\mathrm{res}(\mathbf{r})\,m_\mathrm{eff}\,\mathbf{E}_\mathrm{res}(\mathbf{r})\mathrm{d}V\,,
\end{equation}
so the spatial structure of the modulation does not enter explicitly. Accordingly, the spatially inhomogeneous modulation can be equivalently represented by a constant effective amplitude $m_\mathrm{eff}$. In particular, when the pump is tuned to the frequency of the flat band, the excitation of the high-Q resonance by the pump can be avoided by an appropriate choice of the pump incidence direction and polarization, thereby avoiding resonant field localization inside the nanoresonators that would otherwise lead to strong spatial inhomogeneities in the induced background permittivity modulation.

To be more specific, we study the Kerr effect for the finite bandgap case presented in Fig.~2. The relevant quantity for the Kerr effect is the squared field amplitude $E^2_\mathrm{p}(\mathbf{r})$ of the pump wave. The frequency $\omega_\mathrm{p}$ of the pump wave is identical to $0.5\omega_\mathrm{m}$ used in Figs.~\ref{fig:cylinderqbic}(D)--(E). Moreover, the structural and material parameters of the metasurface are also chosen to be the same as those used in Figs.~\ref{fig:cylinderqbic}(D)--(E). We assume $p$-polarized light incident at an angle $5^{\circ}$ with respect to the axis normal to the metasurface. We point out that, for this polarization, the high-Q modes of the flat band in Fig.~2(B) cannot be excited. The resulting $E^2_\mathrm{p}(\mathbf{r})$ profile inside a cylindrical nanoresonator of the metasurface is shown in Fig.~\ref{fig:S1}. Here, we observe that the pump field distribution inside the nanoresonator is spatially inhomogeneous; however, thanks to the careful choice of the incident polarization, the high-Q mode is not excited. Furthermore, because the modulation amplitude $m$ remains perturbative, the approximation in Eq.~\eqref{eq:kappa2} can be reliably applied.

\section{Comparison of the meta-atom and lattice qBICs}\label{supp-sec:lattice-bic}
\begin{figure*}[h]
\centerline{\includegraphics[width= \columnwidth,trim=0.1 0.1 0.1 0.1,clip]{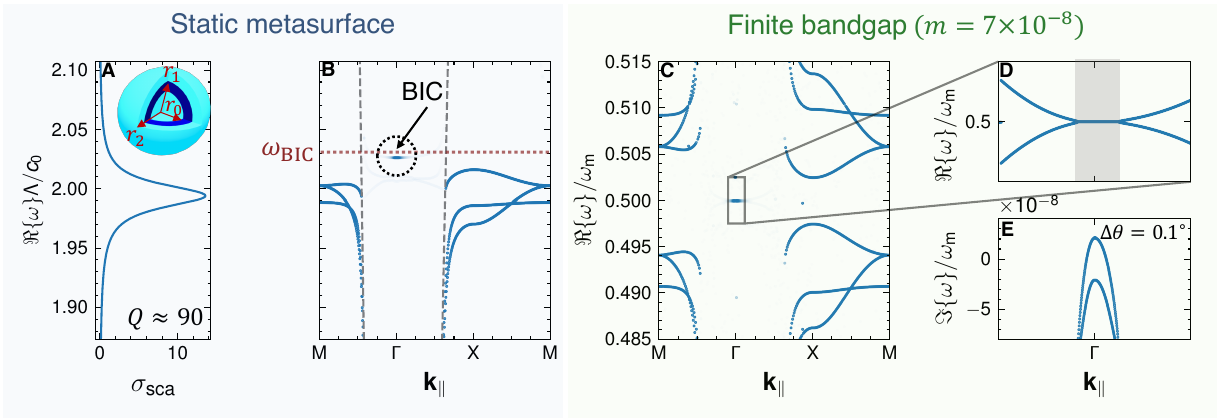}}
\caption{\textbf{PTC assisted by lattice-induced qBIC}. (A) The normalized scattering cross section $\sigma_\mathrm{sca}$ of the static multilayered sphere (see inset) as a function of the frequency $\omega$. The sphere consists of a core and two layers. The core and the outer layer of the sphere are made from AZO. The middle layer is made from silica. 
Note that material losses are neglected here to clearly illustrate the effects. (B) The band structure of the metasurface formed by a square lattice of multilayered spheres under static conditions, {\textit i.e.}, $m=0$ in the spectral domain of interest. The red-dotted line marks the spectral location of the true lattice BIC at the $\Gamma$-point. (C) The band structure of the time-varying metasurface. An extremely low modulation amplitude $m=7\times 10^{-8}$ is used. (D) The further zoomed band structure in the highlighted region of (C) illustrates a finite momentum bandgap. The displayed mode has a symmetry compatible with that of $p$-polarized plane waves. (E) The corresponding imaginary part of the frequency inside the bandgap in (D) for a fixed $\Re{(\omega)/\omega_\mathrm{m}}=0.5$.}
\refstepcounter{SIfig}\label{fig:figS1}
\end{figure*}
\noindent In this section, we show the influence of the qBICs of the metasurfaces formed by lattice interactions on momentum bandgap sizes. We also compare the results obtained thanks to these lattice qBICs with those of the meta-atom qBICs used in the main text. We consider a metasurface made from a square lattice of lossless multilayered spheres (see the inset of Fig.~\ref{fig:figS1}(A)). The core and the outer layer of the spheres are made from lossless AZO. The middle layer is made from lossless silica. To clearly illustrate the impact of the lattice qBICs, the isolated multilayered sphere is designed such that it sustains a relatively low-Q resonance. We use, $r_2=200$~nm, $r_0=66.67$~nm, and $\eta=0.9$. 

The normalized scattering cross section $\sigma_\mathrm{sca}$ of the static multilayered sphere is shown in Fig.~\ref{fig:figS1}(A). Here, we observe that the multilayered sphere hosts an electric dipolar Mie resonance with $Q\approx90$. This Q-factor is three orders of magnitude lower than that of the qBIC resonance used in Fig.~2(D). Next, we plot the band structure of the corresponding static metasurface in Fig.~\ref{fig:figS1}(B). The metasurface is formed by a square lattice of multilayered spheres with lattice period $\Lambda=3r_2$. We observe the existence of flat bands in Fig.~\ref{fig:figS1}(B). However, due to the relatively low Q-factors of the multilayered spheres, the bands are considerably less flat than those shown in Fig.~2(C). This metasurface also supports a true symmetry-protected lattice BIC at the frequency $\omega_\mathrm{BIC}$ at the $\Gamma$-point. Such a BIC is formed due to the lattice interactions in the metasurface \cite{Ustimenko2024resonances}. We note that, despite the presence of a lattice BIC, the flatness of the bands of the metasurface is controlled by the Q-factors of the individual meta-atoms.

To study the impact of the lattice qBICs on the sizes of the momentum bandgaps, we turn on the time modulation of the meta-atoms. We use $m=7\times10^{-8}$, which is identical to that used in Fig.~2(E). Moreover, we choose $0.5\omega_\mathrm{m}\approx\omega'_\mathrm{BIC}$. Here, $\omega'_\mathrm{BIC}$ is the DC-shifted frequency of the lattice BIC. Numerically, $\omega_\mathrm{m}=2\pi\times161$~THz. The band structure of the corresponding time-varying metasurface is shown in Fig.~\ref{fig:figS1}(C). Here, we note that due to the lower flatness of the bands, upon time modulation, a smaller number of modes interact with each other as compared to those in Fig.~2(E) (see the gray box). Therefore, a relatively smaller gap opens despite the same modulation amplitude $m$. We show the corresponding momentum bandgap in Fig.~\ref{fig:figS1}(D). The imaginary part of the frequency $\Im{(\omega)}$ inside the bandgap is shown in Fig.~\ref{fig:figS1}(E). Here, the gap corresponds to $\Delta \theta=0.1^\circ$. This gap size is $40$ times lower than that shown in Figs.~2(F)--(G).

Importantly, the lattice qBICs of the metasurfaces allow for the existence of narrow momentum bandgaps with extremely low $m$. However, it is the qBICs of the individual meta-atoms that assist in expanding the momentum bandgap sizes while simultaneously lowering the modulation amplitude requirements. Physically, the strong localization of the fields inside the individual meta-atoms reduces the spatial dispersion in the lattice while simultaneously allowing for a large interaction time of light with time-varying matter.

\section{Comparison with effectively homogeneous media}\label{supp-sec:homogenous-ptc}
\begin{figure*}[h]
\centerline{\includegraphics[width= \columnwidth,trim=0.1 0.1 0.1 0.1,clip]{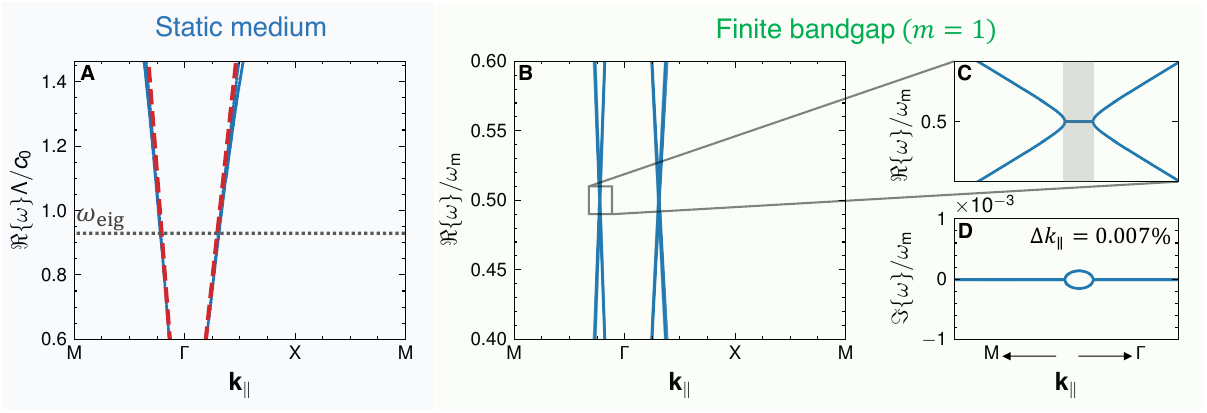}}
\caption{\textbf{Effectively homogeneous medium build based on the metasurface of multilayered spheres}. (A) The band structure of the effectively homogeneous medium under static conditions, {\textit i.e.}, $m=0$. The homogeneous medium is numerically modeled by operating the metasurface in Fig.~2 in its spatially homogenizable regime, \textit{i.e.}, $\Re{(\omega)\Lambda/c_0<\pi}$. The gray-dotted line marks the spectral location of an eigenmode in the M-$\Gamma$ region. Moreover, the red-dashed lines represent the light lines. (B) The band structure of the medium in the time-varying case. A high modulation amplitude $m=1$ is used. (C) The further zoomed band structure in the highlighted region of (B) illustrates a finite momentum bandgap. The displayed mode has a symmetry compatible with that of $p$-polarized plane waves. (D) The corresponding imaginary part of the frequency inside the bandgap in (C) for a fixed $\Re{(\omega)/\omega_\mathrm{m}}=0.5$.}
\refstepcounter{SIfig}\label{fig:sph_slab}
\end{figure*}

\begin{figure*}[h]
\centerline{\includegraphics[width= \columnwidth,trim=0.1 0.1 0.1 0.1,clip]{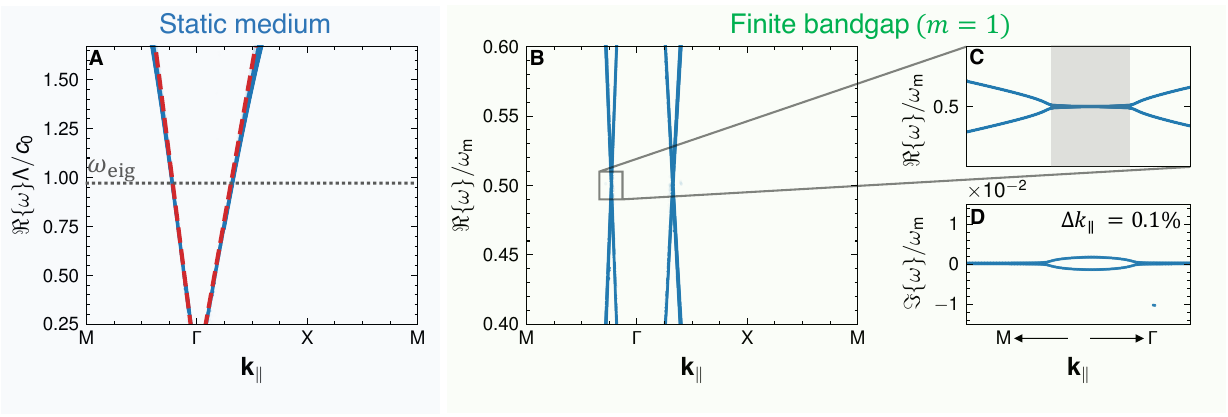}}
\caption{\textbf{Effectively homogeneous medium build based on the metasurface of cylinders}. (A) The band structure of the effectively homogeneous medium under static conditions, {\textit i.e.}, $m=0$. The homogeneous medium is numerically modeled by operating the metasurface in Fig.~4 in its spatially homogenizable regime, \textit{i.e.}, $\Re{(\omega)\Lambda/c_0<\pi}$. The gray-dotted line marks the spectral location of an eigenmode in the M-$\Gamma$ region. Moreover, the red-dashed lines represent the light lines.  (B) The band structure of the medium in the time-varying case. A high modulation amplitude $m=1$ is used. (C) The further zoomed band structure in the highlighted region of (B) illustrates a finite momentum bandgap. The displayed mode has a symmetry compatible with that of $s$-polarized plane waves. (D) The corresponding imaginary part of the frequency inside the bandgap in (C) for a fixed $\Re{(\omega)/\omega_\mathrm{m}}=0.5$.}
\refstepcounter{SIfig}\label{fig:cyl_slab}
\end{figure*}
\noindent In the main text, we showed that by harnessing the flat bands arising due to the qBICs in metasurface-based PTCs, noticeable momentum bandgaps are reached while requiring considerably low modulation amplitudes. Here, we show that in effectively homogeneous media, despite much higher modulation amplitudes, the gap sizes remain rather small. 

To numerically model homogeneous media, we operate the metasurfaces used in the main text in their spatially homogenizable regime. A metasurface is homogenizable if the wavelength $\lambda$ of the considered eigenmode is at least twice its lattice period $\Lambda$, \textit{i.e.}, $\lambda>2\Lambda$. In terms of the eigenmode frequency $\omega$, this condition can be written as $\Re{(\omega)\Lambda/c_0<\pi}$. 

First, we consider the effectively homogeneous medium modeled by operating the metasurface used in Fig.~2 in its spatially homogenizable regime. We choose the same system parameters as those used to plot Fig.~2(C) in the simulations. In particular, the multilayered spheres are assumed to be lossless. The band structure of the static medium in the homogenizable regime is shown in  Fig.~\ref{fig:sph_slab}(A). We observe the characteristic linear dispersion of the homogeneous media. We note that the shown bands appear below the light lines. Here, $\omega_\mathrm{eig}$ is the frequency of the eigenmode in the M-$\Gamma$ region. Next, we choose the modulation frequency $\omega_\mathrm{m}$ such that the linear bands interact upon time modulation. In particular, $0.5\omega_\mathrm{m}=\omega'_\mathrm{eig}$. Here, $\omega'_\mathrm{eig}$ is the DC-shifted frequency of the eigenmode in the M-$\Gamma$ region. Numerically, $\omega_\mathrm{m}=2\pi\times 151$~THz. Furthermore, we use a high modulation amplitude of $m=1$. The band structure in the time-varying case is shown in Fig.~\ref{fig:sph_slab}(B). Here, we observe the crossing of the bands. Due to the linearity of the bands, upon time modulation, very few modes interact with each other as compared to those in Fig.~2(E) (see the gray box). Therefore, an extremely small band gap opens despite the large modulation amplitude. We show the corresponding momentum bandgap in Fig.~\ref{fig:sph_slab}(C). The imaginary part of the frequency $\Im{(\omega)}$ inside the bandgap is shown in Fig.~\ref{fig:sph_slab}(D). To quantify the gap size, we cannot use the angular bandwidth as done in the main text, since the bandgap of the effectively homogeneous medium appears below the light line. Therefore, we define the new gap size metric $\Delta k_\parallel$ as 

\begin{equation}
\label{eq:size}
\Delta k_\parallel=\frac{\delta{k}^\mathrm{M}_{\parallel}+\delta{k}^\mathrm{X}_{\parallel}}{L}\,.
\end{equation}
\noindent
Here, $\delta{k}^\mathrm{M}_{\parallel}$ and $\delta{k}^\mathrm{X}_{\parallel}$ are the momentum extents of the gap in the M-$\Gamma$, and $\Gamma$-X regions of the IBZ, respectively. Moreover, $L$ is the total momentum extent of the M-$\Gamma$-X region, which equals $(\sqrt{2}+1)\frac{\pi}{a}$.

Using the gap size metric in Eq.~\eqref{eq:size}, we find that the bandgap size in Figs.~\ref{fig:sph_slab}(C)--(D) is $\Delta k_\parallel=0.007\%$ for $m=1$. On the other hand, the gap size in Figs.~\ref{fig:sphereMS}(F)--(G) is $\Delta k_\parallel=3\%$ for $m=7\times10^{-8}$. Therefore, we conclude that homogeneous media support a much smaller bandgap, even for huge modulation amplitudes, than the corresponding qBIC-assisted PTCs. 

Next, we tune the cylinder-based metasurface shown in Fig.~4 to its spatially homogenizable regime. Here, we choose the same system parameters as those used to plot Fig.~4(B) in the simulations. The material losses in Ge cylinders are fully considered. The band structure of the medium in the static case is shown in Fig.~\ref{fig:cyl_slab}(A). We again observe characteristic linear dispersion of homogeneous media. Here, $\omega_\mathrm{eig}$ is the frequency of the eigenmode in the M-$\Gamma$ region. The modulation frequency is again chosen such that the linear bands cross upon time modulation. In particular, $0.5\omega_\mathrm{m}=\omega'_\mathrm{eig}$, where $\omega'_\mathrm{eig}$ is the DC-shifted frequency of the eigenmode in the M-$\Gamma$ region. Numerically, $\omega_\mathrm{m}=2\pi\times 29$~THz. Furthermore, a large modulation amplitude of $m=1$ is chosen. The band structure in the time-varying case is shown in Fig.~\ref{fig:cyl_slab}(B). Moreover, the zoomed band structure in the region of band crossing, and the corresponding imaginary part of the frequency $\Im{(\omega)}$ inside the bandgap is shown in Figs.~\ref{fig:sph_slab}(C) and (D), respectively. We observe that for $m=1$, a gap of size $\Delta k_\parallel=0.1\%$ is observed. On the other hand, the gap size in Figs.~\ref{fig:cylinderqbic}(D)--(E) is $\Delta k_\parallel=4\%$ for $m=10^{-4}$. Therefore, we find that homogeneous media support approximately $40$ times smaller bandgap even for $4$ orders of magnitude higher modulation amplitudes than the corresponding qBIC-assisted PTCs.

\section{qBICs of germanium cylinders}\label{supp-sec:static-cylinder-bic}
\begin{figure*}[h]
\centerline{\includegraphics[width= 0.7\columnwidth,trim=0.1 0.1 0.1 0.1,clip]{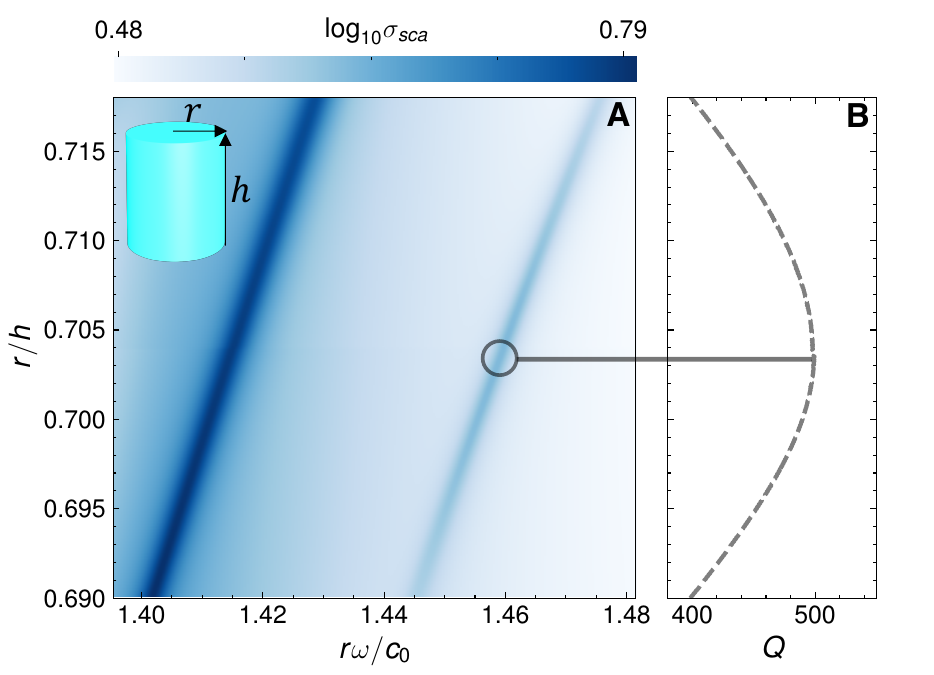}}
\caption{\textbf{The qBIC of the isolated static cylinder}. (A) The rotationally averaged normalized scattering cross section $\sigma_\mathrm{sca}$ of the germanium cylinder (see inset) as a function of the aspect ratio $r/h$ and frequency $\omega$. We change the height $h$ of the cylinder to vary the aspect ratio. The radius $r$ of the cylinder is fixed as $r=1030$~nm. The modes in the right resonance branch represent qBICs formed by the Friedrich-Wintgen mechanism. (B) The quality factor $Q$ of the qBICs in (A) in the right branch as a function of the aspect ratio $r/h$.}
\refstepcounter{SIfig}\label{fig:S2}
\end{figure*}
\noindent In this section, we discuss the qBIC formation mechanism of finite Ge cylinders. This mechanism was first discussed in \cite{rybin2017high}. We plot the rotationally averaged normalized scattering cross section $\sigma_\mathrm{sca}$ of a Ge cylinder as a function of its aspect ratio $r/h$ and frequency $\omega$ in Fig.~\ref{fig:S2}(A). Here, we change $h$ to vary the aspect ratio. The radius $r$ of the cylinder is fixed to $r=1030$~nm (as in the main text). In Fig.~\ref{fig:S2}(A), we observe the emergence of two resonance branches. These branches represent hybrid modes formed by the interaction of Mie-type modes with indices $m=0$, $n=1$ and neighboring Mie-type modes with indices $m=0$, $n=2$. In particular, the far fields of the hybrid modes in the right branch are formed by a destructive interference of the far fields of the parent Mie modes. Due to the destructive interference, these modes possess relatively high Q-factors. In fact, these modes are qBICs formed by the Friedrich-Wintgen mechanism. 

The Q-factors of these qBIC modes in the right branch are plotted in Fig.~\ref{fig:S2}(B). We observe that the maximum Q-factor is achieved for the aspect ratio $r/h\approx0.703$. In the main text, we use Ge cylinders with $r/h\approx0.703$ that host a qBIC with $Q\approx500$. Note that in the limit of vanishing material losses and diverging real part of the permittivity of the cylinders, the Q-factor of the qBIC diverges to infinity, forming a true BIC. 

It is important to mention that the qBIC remains robust against thermal expansion and thermo-optic effects that arise due to optical pumping of Ge. We operate in a frequency regime where Ge has negligible material losses. At the considered pump intensities and using pulsed excitation, the temperature increase $\Delta T$ of the Ge cylinders remains below $1$~K. From Fig.~\ref{fig:S2}, we find that the Q-factor of the qBIC depends on the aspect ratio $r/h$ rather than on the absolute values of $r$ and $h$. Therefore, in the case of isotropic thermal expansion, the aspect ratio and hence the Q-factor of the qBIC is preserved. On the other hand, for anisotropic thermal expansion, using the thermal expansion coefficient of Ge ($6\times10^{-6}\,\mathrm{K}^{-1}$), we find that the maximum relative change in $r$ and $h$ is about $6\times10^{-4}\%$. Therefore, the aspect ratio $r/h$ can not change by more than $10^{-3}\%$. For such small changes in $r/h$, the high Q-factor of the qBIC remains preserved (see Fig.~\ref{fig:S2}(B)).

To assess the impact of the thermo-optic effect, we note that the Q-factor of the qBIC of the cylinders scales as $Q\sim \tilde{n}_\mathrm{r}^{1.6}$ \cite{rybin2017high}, where $\tilde{n}_\mathrm{r}$ is the real part of the refractive index of Ge. Using the thermo-optic coefficient of Ge ($4\times10^{-4}\,\mathrm{K}^{-1}$), at the considered pump intensities, $\tilde{n}_\mathrm{r}$ changes by about $0.01\%$. This variation in $\tilde{n}_\mathrm{r}$ changes the Q-factor of the qBIC by less than $0.02\%$. Therefore, we conclude that the thermo-optic effect does not degrade the Q-factor of qBIC by an appreciable amount.

\section{PTCs based on metasurfaces made from homogeneous germanium spheres}\label{supp-sec:MD-sph}
\begin{figure*}[h]
\centerline{\includegraphics[width= 0.7\columnwidth,trim=0.1 0.1 0.1 0.1,clip]{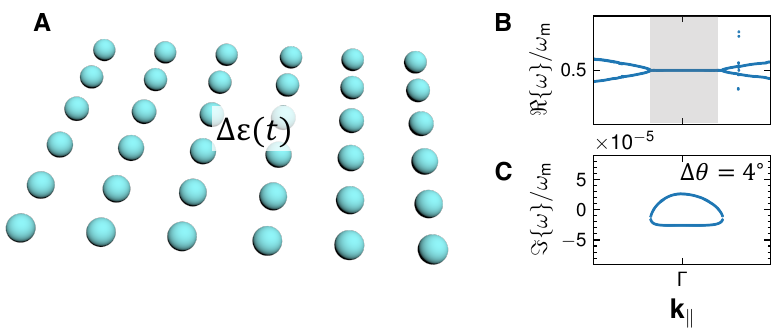}}
\caption{\textbf{PTC assisted by dipolar Mie resonances}. (A) A metasurface-based PTC made from a square lattice of homogeneous spheres supporting the magnetic dipolar Mie resonance. (B) The band structure of the time-varying metasurface made from Mie resonant homogeneous spheres illustrating a finite bandgap that corresponds to $\Delta \theta=4^\circ$. The gap size is identical to that obtained for qBIC resonant cylinder metasurfaces in Fig.~4(D).  Here, $m=1.4\times10^{-3}$ is used, which is an order of magnitude higher than that needed by the cylinder metasurface to open the same-sized bandgap. (C) The corresponding imaginary part of the frequency inside the bandgap, shown in (B), for a fixed $\Re{(\omega)/\omega_\mathrm{m}}=0.5$. The displayed mode branch has a symmetry compatible with that of $p$-polarized plane waves. We use the lattice period $\Lambda=3.5R_0$ for the computations. Here, $R_0$ is the radius of the homogeneous sphere.}
\label{fig:homsphereMS}
\end{figure*}
\noindent A recent study showed that using metasurfaces made from low-loss homogeneous spheres reduces the required modulation amplitude to attain noticeable momentum bandgaps \cite{wang2025expanding}. The study exploits magnetic dipolar resonances of the spheres. We compare our results shown in Figs.~4(D)--(E) for the qBIC resonant cylinder metasurface to those reported in \cite{wang2025expanding}. 

For a fair comparison, we consider a metasurface composed of a square lattice of homogeneous Ge spheres (see Fig.~\ref{fig:homsphereMS}(A)). The radius $R_0$ of the spheres is optimized such that the flat bands of the metasurface due to the magnetic dipolar resonance occur at the frequency $\omega=2\pi\times67.3$~THz. Here, the frequency of the flat bands of the spheres is close to that of the cylinders. This choice ensures that the homogeneous sphere metasurface experiences a similar material dispersion as that of the cylinder metasurface at the spectral location of the momentum bandgap. Therefore, we use $\omega_\mathrm{m}=2\pi\times134.6$~THz. The optimized radius of the spheres is $R_0= 550$~nm. Furthermore, the quality factor of the magnetic dipolar resonance of the optimized homogeneous static sphere is $Q\approx15$.

We plot the band structure of the time-varying metasurface in Fig.~\ref{fig:homsphereMS}(B). Moreover, we also show the imaginary part of the frequency inside the momentum bandgap (see Fig.~\ref{fig:homsphereMS}(C)). We find that to open a bandgap of size $\Delta \theta=4^\circ$, a modulation amplitude of $m=1.4\times 10^{-3}$ is required. This value of $m$ corresponds to the pump intensity of $9.8$~$\mathrm{GW/cm^2}$, which is too close to the damage threshold of structured Ge using multi-shot pulses \cite{malik2013determination}. On the other hand, to open the same-sized bandgap for the metasurface with cylinders, $m= 10^{-4}$ is needed, which is an order of magnitude below the damage threshold of structured Ge. Furthermore, the cylinder-based geometry is much easier to fabricate compared with the sphere-based metasurface.

\section{Link between the T-matrix and S-matrix formalism}\label{supp-sec:smat}
The S-matrix $\mathbf{\hat{S}}_\mathrm{0}$ of an isolated time-varying scatterer can be calculated from its T-matrix $\mathbf{\hat{T}}_\mathrm{0}$ as \cite{waterman2009tmatrix}
\begin{equation}
\mathbf{\hat{S}}_\mathrm{0}=\mathbf{\hat{I}}+2\mathbf{\hat{T}}_\mathrm{0}\,.\label{eq:Smatsphere}
\end{equation}
Furthermore, for a metasurface made from such scatterers with the effective T-matrix $\mathbf{\hat{T}}_\mathrm{eff}$, the corresponding S-matrix $\mathbf{\hat{S}}$ is given by

\begin{equation}
\label{eq:Teff_c}
\mathbf{\hat{S}}=\mathbf{\hat{I}}+\mathbf{\hat{V}}\hspace{2pt}\mathbf{\hat{T}_\mathrm{eff}}\hspace{2pt}\mathbf{\hat{U}}\,.
\end{equation}
\noindent
Here, $\mathbf{\hat{V}}$ is the coordinate transformation matrix of the scattered field from the spherical wave to the plane wave basis, and $\mathbf{\hat{U}}$ is the coordinate transformation matrix of the incident field from the plane wave to the spherical wave basis \cite{garg2022modeling}.

\section{Details of the convergence parameters used during simulations}\label{supp-sec:convergence-params}
For performing the simulations throughout the article, we used various convergence parameters. These parameters are the maximum multipolar order $l_\mathrm{max}$, and the total number of scattered frequencies $J$. Here, we list the numerical values of these parameters.

\noindent Figures~2(A),(C),(D): $l_\mathrm{max}=4,\,\textrm{and }J=1$.\\
\noindent Figures~2(E)--(H): $l_\mathrm{max}=4,\,\textrm{and }J=7$.\\
\noindent Figures~3(A)--(B): $l_\mathrm{max}=4,\,\textrm{and }J=7$.\\
\noindent Figures~4(B)--(C): $l_\mathrm{cut}=10$, $l_\mathrm{max}=6,\,\textrm{and }J=1$.\\
\noindent Figures~4(D)--(F): $l_\mathrm{cut}=10$, $l_\mathrm{max}=6,\,\textrm{and }J=7$.\\
\noindent Figures~S2(A)--(B): $l_\mathrm{max}=4,\,\textrm{and }J=1$.\\
\noindent Figures~S2(C)--(E): $l_\mathrm{max}=4,\,\textrm{and }J=7$.\\
\noindent Figures~S3(A): $l_\mathrm{max}=4,\,\textrm{and }J=1$.\\
\noindent Figures~S3(B)--(D): $l_\mathrm{max}=4,\,\textrm{and }J=7$.\\
\noindent Figures~S4(A): $l_\mathrm{cut}=6\,,l_\mathrm{max}=4,\,\textrm{and }J=1$.\\
\noindent Figures~S4(B)--(D): $l_\mathrm{cut}=6\,,l_\mathrm{max}=4,\,\textrm{and }J=7$.\\
\noindent Figures~S5(A)--(B): $l_\mathrm{cut}=10\,,l_\mathrm{max}=6,\,\textrm{and }J=1$.\\
\noindent Figures~S6(B)--(C): $l_\mathrm{max}=4,\,\textrm{and }J=7$.\\

%
\end{document}